\documentclass[aps,pra,twocolumn,tightenlines,floatfix,showpacs]{revtex4}
\usepackage[dvips]{graphicx}
\usepackage[english]{babel}
\usepackage{amsmath}
\usepackage{amssymb}
\usepackage{times,bm}
\newcommand{\bs}{\begin{split}}
\newcommand{\es}{\end{split}}
\newcommand{\bea}{\begin{eqnarray}}
\newcommand{\eea}{\end{eqnarray}}
\newcommand{\be}{\begin{equation}}
\newcommand{\ee}{\end{equation}}
\newcommand{\ba}{\begin{eqnarray}}
\newcommand{\ea}{\end{eqnarray}}
\newcommand{\ek}{\epsilon_{\mathbf{k}}}
\newcommand{\Ek}{E_{\mathbf{k}}}

\newcommand{\vk}{v_{\mathbf{k}}}
\newcommand{\xik}{\xi_{\mathbf{k}}}

\begin{document}

\title{Theory of Superfluids with Population Imbalance:
Finite Temperature and BCS-BEC Crossover Effects}

\author{Qijin Chen, Yan He, Chih-Chun Chien and K. Levin}

\affiliation{James Franck Institute and Department of Physics,
 University of Chicago, Chicago, Illinois 60637}

\date{\today}

\begin{abstract}
In this paper we present a very general theoretical framework for
addressing fermionic superfluids over the entire range of BCS to Bose
Einstein condensation (BEC) crossover in the presence of population
imbalance or spin polarization.  Our emphasis is on providing a theory
which reduces to the standard zero temperature mean field theories in
the literature, but necessarily includes pairing fluctuation effects at
non-zero temperature within a consistent framework. Physically, these
effects are associated with the presence of pre-formed pairs (or a
fermionic pseudogap) in the normal phase, and pair excitations of the
condensate, in the superfluid phase. We show how this finite $T$ theory
of fermionic pair condensates bears many similarities to the
condensation of point bosons.  In the process we examine three different
types of condensate: the usual breached pair or Sarma phase and both the
one and two plane wave Larkin- Ovchinnikov, Fulde-Ferrell (LOFF) states.
The last of these has been discussed in the literature albeit only
within a Landau-Ginzburg formalism, generally valid near $T_c$. Here we
show how to arrive at the two plane wave LOFF state in the ground state
as well as at general temperature $T$.
\end{abstract}


\maketitle

\section{Introduction}
\label{sec:I}
The subject of superfluidity in ultracold trapped fermionic gases is an
exciting field
\cite{Jin3,Grimm,Jin4,Ketterle3,KetterleV,Thomas2,Grimm3,ThermoScience},
not only for its implications in atomic physics, but also because there
are important implications for condensed matter systems, including
perhaps high temperature superconductors \cite{ourreview,ReviewJLTP}.
There are two important aspects which are particularly notable about
these systems, from the perspective of the present paper. They can be
tuned in various ways which are not available to nature-made
superconductors.  Thus, one can study the entire regime from BCS to
Bose-Einstein condensation (BEC), simply by the application of a
magnetic field in concert with a Feshbach resonance.  Moreover, one can
vary the concentrations of the two spin species arbitrarily
\cite{ZSSK06,PLKLH06,ZSSK206}; in a fashion, this simulates the
application of a magnetic Zeeman field.  This latter tuneability has
important implications for other sub-disciplines in physics such as
dense QCD and (isospin asymmetric) nuclear matter
\cite{Wilczek,LW03,FGLW05}.  Equally important is the fact that there is
a rich collection of experimental data from two different atomic physics
groups \cite{ZSSK06,PLKLH06,ZSSK206} on $^6$Li gases near unitarity,
with which one can compare various theoretical results.

The goal of this paper is to present an overarching theoretical
framework for dealing with fermionic gases over the entire range of BCS
to BEC crossover and the entire range of temperature, as well as the
entire range of population imbalance.  The foundations of this theory
lie with the initial observation of Eagles \cite{Eagles} and of Leggett
\cite{Leggett} that the BCS-like wavefunction has a much greater
generality than was originally recognized at the time of its
proposal. It is capabable of describing both BCS and BEC like systems,
providing the pairing attraction is tuneable from arbitrarily weak (BCS)
to arbitrarily strong (BEC) and one self consistently solves for the
fermionic chemical potential. This mean field-like ground state wave
function is also readily generalized to include population
balance. Indeed, there are at least three well studied phases
\cite{Sarma63,Caldas04,FFLO} which have been proposed to accomodate a
difference in the population of the two spin species.  What we want to
stress is that these same mean field theories have a natural extension
to finite temperature \cite{ourreview}.  This extension will be a focus
of the present paper.

At strictly zero temperature, there is a rather extensive literature
\cite{SR06,SM06,HS06, Kinnunen,Tsinghuagroup, PWY05,PS05a} on these
population-imbalanced superfluids and superconductors, including
recently the effects of the crossover from BCS to BEC.  There have also
been some studies at finite $T$, which are at the mean field level and
do not include the effect of the non-condensed pairs
\cite{Machida2,YD05,LatestStoof} we consider here.  
The three most studied phases are the
``Sarma"-like \cite{Sarma63} or breached pair states in which (as in BCS
theory) the condensed pairs have zero net momentum, but polarization
can, nevertheless, be introduced. At $T=0$ this state appears to be
stable in the deep BEC regime.
Additionally, two different
phases \cite{FFLO} have been proposed by Larkin and Ovchinnikov and by
Fulde and Ferrell (LOFF) in which the condensate has a net momentum of a
pre-determined $\mathbf{q}$ or of $\pm \mathbf{q}$.  Even more elaborate
crystalline lattices of various $\mathbf{q}_i$ have also been
contemplated \cite{LOFF_Review}.  It is believed that these states are
more appropriate closer to the BCS side of resonance, although rather
little has been determined about the ``two-plane wave" LOFF state near
$T=0$ or in the presence of crossover effects.  Added to this complexity
is the possiblity of heterogeneous or phase separated states
\cite{Caldas04}.

In this paper we will present the theoretical formalism for the Sarma
and one and two-plane wave LOFF states at zero and finite $T$ as one
varies from BCS to BEC.  We note that because theories of population
imbalanced superfluids are (currently, without exception) based on
BCS-Leggett-type \cite{Leggett} ground states, it is important to
determine their finite temperature implications within this broad class
of ground states, as we do here.  Our premise is that the effects of
finite $T$, which necessarily must be accomodated in any comparison with
experiment, must be compatible with the $T=0$ formalism.  Indeed, one of
the most important effects of temperature is to stabilize the Sarma-like
phase.  In this way one finds an intermediate temperature superfluid
\cite{Chien06,ChienRapid,Sedrakian,Stability}, one that exists at $T
\neq 0$, but not at $T=0$.  Our studies of the two-plane wave LOFF state
present new results by extending the current literature away from the
Landau-Ginzburg regime (near $T_c$).

With this class of generalized mean field theories we will show that the
effects of temperature enter in a fashion which is strongly reminiscent
of Bose gas condensation.  Except in the BCS limit, pairs form at higher
temperatures, than the (transition) temperature $T_c$ at which they
condense.  Thus we have to distinguish the excitation gap [called
$\Delta(T)$] from the order parameter [called
$\Delta_{sc}(T)$]. Alternatively, this means that there is an excitation
gap (or pseudogap $\Delta_{pg}^2$) for fermionic excitations even in the
normal phase.  It also implies that below $T_c$ there will be additional
modes of exciting the condensate-- via pair excitations. These latter
are non-condensed or incoherent pairs with finite center of mass
momentum. We have found \cite{ChienRapid} that in a trapped geometry
they are particularly important for providing a mechanism of getting
polarization into the gas.

\begin{table*}[!bht]
\begin{center}
\begin{tabular}{|p{1.5in}|p{2.2in}|p{1.7in}|}
\hline
& \parbox[c][7mm][c]{2.2in}{Composite bosons} &
\parbox[c][7mm][c]{1.7in}{Point bosons} \\ 
\hline
\centering{Pair chemical potential} &
\parbox[c][10mm][c]{2.2in}{$\mu_{pair}=0$, 
  $T\leq T_c$ \\ Leads to BCS gap equation for $\Delta(T)$} &
\centerline{$\mu_{B}=0$, $T\leq T_c$} \\ 
\hline
\parbox[c][8mm][c]{1.5in}{Total ``number'' of pairs} &
\parbox[c][8mm][c]{2.2in}
{$\Delta^{2}(T)=\Delta_{sc}^{2}(T)+\Delta_{pg}^{2}(T)$}
& \parbox[c][8mm][c]{1.7in}{$N=N_{0}+N_{T}$} \\ 
\hline
\parbox[c][8mm][c]{1.5in}{Noncondensed pairs} &
\parbox[c][8mm][c]{2.2in}{$Z\Delta_{pg}^{2}=\sum_{\mathbf{q}\neq
    0}b(\Omega_{q})$} & 
\parbox[c][8mm][c]{1.7in}{$N_{T}=\sum_{\mathbf{q}\neq 0}b(\Omega_{q})$} \\
\hline
\end{tabular}
\caption{BCS theory by way of BEC analogy.  Here we compare condensation
in composite and point bosons; $\mu_B$ is the bosonic chemical
potential, $N_0$ is the number of condensed and $N_T$ is the number of
non-condensed bosons. We define $\mu_{pair}$ as the chemical potential
for the non-condensed pairs.  Here $\Delta(T)$ is the total fermionic
gap which contains contributions from the non-condensed
($\Delta_{pg}^2$) and condensed terms ($\Delta_{sc}^2$). In the strict
BCS limit $\Delta_{pg} =0$, so that the order parameter and gap are
identical. }
\end{center}
\end{table*}

These effects of finite $T$ can be compared with an alternative class of
theories in the literature based on work by Nozieres and Schmitt-Rink
(NSR) \cite{NSR}. This approach is known to lack self-consistency
\cite{Serene}. We stress that the finite temperature NSR approach was
not designed to be consistent with the standard ground state equations.
This observation has also been made in Ref. \cite{Strinati5}. Indeed,
these latter authors have presented in considerable detail \cite{PS05} a
more complete finite $T$ approach based on the $T_c$ calculations of
Ref.~\cite{NSR}.  A major concern about this class of theories remains
to be addressed.  Because the gap and the number equations are not
treated on an equivalent basis, it is possible that the superfluid
density will not consistently vanish at $T_c$. At this temperature there
has to be a precise, but delicate cancellation of paramagnetic and
diamagnetic current contributions, as found in the present theory
\cite{Kosztin2,JS}.  In a related fashion, pseudogap effects (associated
with the presence of non-condensed pairs) appear, within an NSR-based
approach, in the number equation but not in the gap equation.  

We begin at the more physical level by stressing the analogy between
condensation in this composite boson or fermionic superfluid and
condensation in a gas of ideal point bosons. Our theory treats
self-consistently two-particle and one-particle Green's functions on an
equal footing.  Because the physics is so simple and clear, we can
fairly readily anticipate the form of the central equations of this
BCS-BEC generalization of BCS theory. It is important to stress,
however, that these equations can be derived more rigorously from a
truncated series of equations of motion for the appropriate
Green's functions
\cite{Chen2}.

There are three principle equations which govern Bose condensation: the
vanishing of the bosonic chemical potential at all $ T \leq T_c$ is the
first.  Throughout this paper we will refer to this condition as the
``BEC condition".  It is related to the usual Thouless criterion, but
the latter is generally associated only with the temperature $T_c$. The
second equation is the boson number equation.  All ``bosons" must be
accounted for as either condensed or non-condensed.  The third equation
is the number of non-condensed ``bosons", which are created by thermal
excitations. This is determined simply by inserting the known excitation
spectrum of the excited pairs or bosons, into the Bose distribution
function.  With this equation, and the first equation, one can then
deduce the number of condensed bosons.

These three central equations for bosons are indicated in Table I, on
the far right, for true point bosons, and in the second column for the
composite bosons which appear in fermionic superfluids.
For these composite bosons the quantity which provides a measure of the
``number" of bosons ($N$) is given by $\Delta^2(T)$ (up to a constant
coefficient, $Z$).  This is reasonably easy to see.  In the fermionic
regime, when the fermionic chemical potential is positive, $\Delta^2(T)$
represents the square of the excitation gap.  This is the energy which
must be supplied to break apart the pairs.  Thus, 
$\Delta^2(T)$, in some sense then, reflects the number of pairs.  How
does one quantitatively establish the appropriate ``boson number" for
the fermionic case?  This is determined via the self consistent gap
equation for $\Delta(T)$, which, in turn, is determined using the first
condition: that the pair chemical potential is zero at and below $T_c$.
How does one compute the number of excited pairs?  Once the gap equation
is interpreted in terms of the appropriate non-condensed pair
propagator, then one knows the related excitation spectrum $\Omega_q$ of
this propagator.  

The quantity $Z$ which appears in the last equation of the Table (for
the composite bosons) gives the relation between the gap associated with
non condensed pairs ($\Delta_{pg}^2$) and the number of pairs ($\sum
b(\Omega_q)$). It can be readily calculated in this theory; once one has
the non-condensed pair propagator, $Z$ appears as the inverse residue.
(Deep in the BEC regime, $Z$ is relatively simple to compute, for here
the boson number density approaches the asymptote $n/2$, where $n$ is
the fermion density).
More precisely, the total number of bosons in the present case has to be
determined self-consistently through the gap equation. It also involves
the fermion number equation through the related fermionic chemical
potential.  In this last context, it should be stressed that there is
one important aspect of the fermionic superfluids, which is not apparent
in Table I.  For BCS-BEC crossover, it is essential to derive the self
consistent equation for the fermionic chemical potential; in this
problem the fermions are the fundamental statistical entity. This can be
readily accomplished within the same framework used to arrive at the gap
equation. The vanishing of the pair chemical potential is associated
with a particular choice for the pair propagator involving dressed
Green's functions.  These, in turn, determine the fermionic chemical
potential through the fermion number equation.  In the next two sections
we turn to the gap and number equations, and show through a Green's
function formulation, how strongly these two equations are
inter-connected.

The rest of the paper is organized as follows. We conclude this
section with a summary of the central equations associated
with our $T$-matrix scheme. In Sec. \ref{sec:II}, we
present a mean-field theory for the Sarma and one-plane-wave LOFF state
for $T<T_c$. Due to the complexity of the two-plane-wave LOFF state, we
dedicate an entire section (Sec.\ref{sec:III}) to its
mean-field treatment. In Sec. \ref{sec:IV} we 
present a generalization of our $T$-matrix formalism to include
pairing fluctuation effects. Section \ref{sec:V}
recapitulates our simple physical picture and Section
\ref{sec:VI} presents our conclusions. 
Additional, more technical details are given in
two Appendices for the one and two plane wave LOFF states.

\subsection{Central Equations of $T$-matrix scheme}
\label{sec:Ia}

To make contact with the general class of mean field theories (including
the Sarma and LOFF states), we introduce a $T$-matrix approximation.
This means that we consider the coupled equations between the particles
(with propagator $G$) and the pairs [with propagator $t(P)$] and drop
all higher terms.  This theory does not include direct ``boson-boson"
interactions, although the pairs do interact indirectly via the
fermions, in an averaged or mean field sense.
Throughout Sections \ref{sec:II} and \ref{sec:III} of this paper we will
be showing that the BEC condition noted in the previous section will
give the same gap equation we find using standard techniques, such as
Bogoliubov diagonalization applied to the linearized mean field
Hamiltonian.  Here, for all $T\leq T_{c}$, the BEC condition is
interpreted as requiring that the pair chemical potential $\mu_{pair}$
associated with the non-condensed pairs vanish.  In Sections
\ref{sec:II} and \ref{sec:III} we will address only the first line of
Table I.  The second two lines, or sets of equations will be discussed
in Section \ref{sec:IV}.

Within a $T$-matrix scheme, the pair propagator is given by
\begin{equation}
t^{-1}(P)=U^{-1}+\chi(P)
\label{eq:1}
\end{equation}
where $\chi$ is the spin \emph{symmetrized} pair susceptibility, and
$U<0$ is the pairing interaction strength.  The function $\chi(P)$ is,
in many ways, the most fundamental quantity we introduce in this paper.
It provides the basis for obtaining well known (as well as new) results
of the zero temperature theory. Moreover, it provides the basis for
arriving at a finite temperature description, which appears in Section
\ref{sec:IV}.  The introduction of spin symmetrization is only important
for the case of population imbalance.  In earlier literature
\cite{ourreview}, this complexity did not arise.  We will show that one
obtains consistent answers between $T$-matrix based approaches and
standard mean field theories, provided the components of the pair
susceptibility in the presence of population imbalance, defined by
\begin{equation}
\chi(P)=\frac{1}{2}\big[\chi_{\uparrow\downarrow}(P)+
\chi_{\downarrow\uparrow}(P)\big]
\label{eq:2}
\end{equation}
is given by the product of one dressed and one bare Green's function

\begin{subequations}
\label{eq:3}
\begin{eqnarray}
\label{eq:3a}
\chi_{\uparrow\downarrow}(P)&=&\sum_{K}G_{0\uparrow}(P-K)G_{\downarrow}(K)  \\
\chi_{\downarrow\uparrow}(P)&=&\sum_{K}G_{0\downarrow}(P-K)G_{\uparrow}(K)
\label{eq:3b}
\end{eqnarray}
\end{subequations}
%
where $P=(i\Omega_l,\mathbf{p})$, and $G$ and $G_0$ are the full and
bare Green's functions respectively. We will discuss $G$ in more detail
on a case by case basis.  Here $G_{0,\sigma}^{-1} (K) = i \omega_{n} -
\xi_{\mathbf{k},\sigma}$, $\xi_{\mathbf{k},\sigma} = \ek-\mu_\sigma$,
$\ek=\hbar^2k^2/2m$ is the kinetic energy of fermions, and $\mu_\sigma$
is the fermionic chemical potential for spin
$\sigma=\uparrow,\downarrow$.  Throughout this paper, we take $\hbar=1$,
$k_B=1$, and use the four-vector notation $K\equiv (i\omega_n,
\mathbf{k})$, $P\equiv (i\Omega_l, \mathbf{q})$, $\sum_K \equiv T\sum_n
\sum_{\bf k}$, etc, where $\omega_n = (2n+1)\pi T$ and $\Omega_l = 2l\pi
T$ are the standard odd and even Matsubara frequencies \cite{Fetter}
(where $n$ and $l$ are intergers).

For the mean field discussions in Sections \ref{sec:II} and
\ref{sec:III} of this paper we will not be considering general values of
$P$ but only zero frequency limits with special values of $\mathbf{p}$
associated with Sarma ($\mathbf{p}=0$) and LOFF ($\mathbf{p}=\mathbf{q}
\ne 0$) states. However, when we include the contribution of
non-condensed pairs (or pseudogap effects) the general values of $P$
become important.
For the Sarma phase we have the BEC condition 
\begin{equation}
t^{-1} (0) = 0 = U^{-1}+\chi(0)
\label{eq:5}
\end{equation}
%
%
More generally, for LOFF-like states we have the BEC condition
at finite $\mathbf{q}$:
\begin{equation}
U^{-1}+\chi(0,\mathbf{q})=0
\label{eq:7}
\end{equation}
We will discuss, in considerable detail, the nature of the mean field
self energy which appears in the full Green's function $G_\sigma(K)$.
Not only does this determine the gap equation but it also leads to the
number equations which can be written in terms of Green's functions as
$n_{\sigma}=\sum_{K}G_{\sigma}(K)$.

The number and gap equations then provide the underlying
basis for the mean field approach.  And we will see that the
\textit{same} propagator for non-condensed pairs [$t(P)$] enters into
the beyond-mean-field corrections.
It is important to stress that when we refer to "mean field" based
approaches in sections \ref{sec:II} and \ref{sec:III} of this paper, we
will not be distinguishing between the order parameter $\Delta_{sc}$ and
the excitation gap $\Delta$.  Subsequently we will show that this
distinction is actually an important one in all but the BCS
limit. Specifically, we note that the expressions we present in Sections
\ref{sec:II} and \ref{sec:III} within our Green's function-based
formulation, are more generally valid below $T_c$, but for the
excitation gap $\Delta$, not for the order parameter.

\section{Gap and Number Equations of Sarma and One Plane
Wave LOFF Phases}
\label{sec:II}

\subsection{Mean Field Sarma State}

We now want to study the Sarma or breached pair state.  On the fermionic
side of resonance, the dispersion relation of one of the two
quasi-particle bands ($E_{k,\uparrow}$ and $E_{k,\downarrow}$, defined
below) has two zero crossings, at momenta $k_1$ $(\ge 0)$ and $k_2$ $ (>
k_1)$.  This is associated with gapless excitations.  In a momentum
space representation, pairing is confined to $k < k_1$ and $k > k_2$.
This
pairing is "breached" by a normal component in the intermediate region
$k_1 < k < k_2$.  It is this normal component, then, which carries the
bulk of the polarization.

If we consider pairing between $\mathbf{k}$ and $-\mathbf{k}$
states, the mean field Hamitonian can be rewritten as
\begin{eqnarray}
H^{MF}&=&\sum_{\mathbf{k}}\Big\{\xi_{\mathbf{k},\uparrow} 
c_{\mathbf{k},\uparrow}^{\dag}c_{\mathbf{k},\uparrow}
+\xi_{\mathbf{k},\downarrow} c_{\mathbf{-k},\downarrow}^{\dag}
c_{\mathbf{-k},\downarrow}
\nonumber \\
& &{}+\Delta c_{\mathbf{-k},\downarrow}^{\dag}c_{\mathbf{k},\uparrow}^{\dag}+
\Delta
c_{\mathbf{k},\uparrow}c_{\mathbf{-k},\downarrow}\Big\}\,,
\label{eq:8}
\end{eqnarray}
where we have chosen $\Delta^* = \Delta$ to be real.  Using standard
Bogoliubov diagonalization techniques, we readily arrive at the mean
field gap equation via the self-consistency condition
\begin{equation}
\Delta \equiv\sum_{K} U \langle
c_{k,\uparrow}c_{-k,\downarrow}\rangle \,,
\label{eq:9}
\end{equation}
which can be written as
\begin{eqnarray}
0 
&=&\frac{1}{U}+\sum_{\mathbf{k}}
  \left[\frac{1-f(E_{k\downarrow})-f(E_{k\uparrow})}{2E_{k}}  
  \right]\nonumber\\
&=& \frac{1}{U}+\sum_{\mathbf{k}}\frac{1-2\bar{f}(E_{k})}{2E_{k}} \,.
\label{eq:10}
\end{eqnarray}
%
Here $\mu=(\mu_{\uparrow}+\mu_{\downarrow})/2$ and
$h=(\mu_{\uparrow} -\mu_{\downarrow})/2$, 
$\Ek = \sqrt{\xik^2 +\Delta^2}$, 
$E_{k\uparrow}=-h+E_{k}$ and $E_{k\downarrow}=h+E_{k}$, where
$\xi_{k}=\epsilon_{k}-\mu$.
In addition, we define the average $\bar{f}(x) \equiv
[f(x+h)+f(x-h)]/2$, where $f(x)$ is the Fermi distribution function. The
coupling constant $U$ can be replaced in favor of the dimensionless
parameter, $1/k_Fa$, via the relationship $m/(4\pi a) = 1/U +
\sum_{\mathbf{k}}(2\epsilon_{k})^{-1}$, where $a$ is the two-body
$s$-wave scattering length, and $k_F$ is the noninteracting Fermi wave
vector for the same total number density in the absence of population
imbalance.
Therefore the gap equation can be rewriten as
\begin{equation}
-\frac{m}{4\pi a}=\sum_{\mathbf{k}}\left[\frac{1-2\bar{f}(E_{k})}
{2E_{k}}-\frac{1}{2\epsilon_{k}} \right] \,.
\label{eq:11}
\end{equation}
This is a familiar gap equation which has appeared many times in the
literature.
The mean field number equations can be readily deduced 
\begin{eqnarray}
n_{\sigma}&=&\sum_{\mathbf{k}}[f(E_{k\sigma})u_{\mathbf{k}}^{2}+
f(E_{k\bar{\sigma}}) v_{\mathbf{k}}^{2}] \,,
\label{eq:12}
\end{eqnarray}
where $\bar{\sigma} = -\sigma$ and the coherence
factors $u_\mathbf{k}^2, v_\mathbf{k}^2 = (1\pm
\xi_\mathbf{k}/\Ek)/2$. 
Equivalently, they can be rewritten as 
\begin{subequations}
\label{eq:neq}
\begin{eqnarray}
\label{eq:neqa}
n &=& 2\sum_\mathbf{k} \left[\vk^2 + \frac{\xi_\mathbf{k}}{\Ek}
  \bar{f}(\Ek)\right],\\
pn &=& \sum_\mathbf{k} [f(\Ek-h)-f(\Ek+h)] \,,
\label{eq:neqb}
\end{eqnarray}
\end{subequations}
where $n= n_\uparrow +n_\downarrow$ is the total atomic density, $\delta
n = n_\uparrow -n_\downarrow >0$ is the number difference and $p=\delta
n/n$ is the polarization.


\subsection{ $T$-matrix Approach Below $T_c$: Sarma State}
\label{sec:2b}

We next show that the same results can be obtained from a $T$-matrix
based approach.  This discussion provides the link between the first
line of Table I and the previous subsection. The one particle Green's
function for particles with spin $\sigma$ is
\begin{eqnarray}
G^{-1}_{\sigma}(K)&=&G^{-1}_{0\sigma}(K)-\Sigma_{\sigma}(K) \nonumber \\
&=& i\omega_{n} -\xi_{k\sigma}-\Sigma_{\sigma}(K) \,,
\label{eq:13}
\end{eqnarray}
%
The self-energy $\Sigma_{\sigma}$ can be shown to be of the BCS-like
form
\begin{equation}
\Sigma_{\sigma}(K)=-\Delta^{2}G_{0\bar{\sigma}}(-K) = 
\frac{\Delta^{2}}{i\omega+\xi_{k\bar{\sigma}}} \,.
\label{eq:14}
\end{equation}
We will see later in Section \ref{sec:IV} how this form for the self
energy very naturally arises (below $T_c$) in a $T$-matrix approach.
Thus
\begin{equation}
G^{-1}_{\sigma}(K)= i\omega-\xi_{k\sigma}-\frac{\Delta^{2}}
{i\omega+\xi_{k\bar{\sigma}}}\,. 
\label{eq:15}
\end{equation}
Then, using the coherence factors we defined earlier,
the Green's functions become
\begin{equation}
G_{\sigma}(K)=\frac{u_{k}^{2}}{i\omega-E_{k\sigma}} +
\frac{v_{k}^{2}}{i\omega+E_{k\bar{\sigma}}} \,.  
\label{eq:16}
\end{equation}

Now we are in position to calculate the pair susceptibility at $P=0$ for
the Sarma phase based on Eqs. (\ref{eq:2}) and (\ref{eq:3}).
\begin{eqnarray}
\chi(0) &=& \chi_{\uparrow\downarrow}(0) =
\chi_{\downarrow\uparrow}(0)\nonumber\\ 
&=&-\sum_{K}\frac{1}{(i\omega_{n}-E_{k\downarrow})(i\omega_{n}+E_{k\uparrow})}
\,.
\label{eq:17}
\end{eqnarray}
%
%
Substituting this expression into our BEC condition Eq.~(\ref{eq:5}), we
 obtain the same gap equation (\ref{eq:10}), after carrying out the
 Matsubara summation.

%
%
In terms of Green's functions, we readily arrive at the number
equations: $n_{\sigma}=\sum_{K}G_{\sigma}(K)$, which reduce to the
number equations (\ref{eq:12}) we found earlier.

\subsection{Mean Field Theory of One Plane Wave LOFF State}
\label{sec:IIB}

If we now consider condensates in which momentum $\mathbf{k}$ pairs with
$\mathbf{-k+q}$, for, as yet undermined $\mathbf{q}$, the mean field
Hamiltonian can be rewritten as
\begin{eqnarray}
H^{MF}&=&\sum_{\mathbf{k}} \Big\{\xi_{\mathbf{k},\uparrow}
c_{\mathbf{k},\uparrow}^{\dag}c_{\mathbf{k},\uparrow} 
+\xi_{\mathbf{k-q},\downarrow}
c_{\mathbf{-k+q},\downarrow}^{\dag}c_{\mathbf{-k+q},\downarrow} 
 \nonumber \\
& &{}+\Delta c_{\mathbf{-k+q},\downarrow}^{\dag}
 c_{\mathbf{k},\uparrow}^{\dag}+\Delta
 c_{\mathbf{k},\uparrow}c_{\mathbf{-k+q},\downarrow}\Big\} 
\label{eq:19}
\end{eqnarray}
%
%
Upon Bogoliubov diagonalization, the self-consistency condition
\begin{equation}
\Delta \equiv \sum_{K} U \langle
c_{k,\uparrow}c_{-k+q,\downarrow}\rangle
\label{eq:20}
\end{equation}
%
readily leads to the mean-field gap equation
\begin{eqnarray}
0&=&\frac{1}{U}+\sum_{\mathbf{k}} \frac{1-f(E_{1,\downarrow})-
f(E_{1,\uparrow})}{2E_{kq}}
\,,
\label{eq:21}
\end{eqnarray}
with $E_{1,\uparrow}=E_{kq}-h+(\epsilon_k-\epsilon_{k-q})/2$,
$E_{1,\downarrow}=E_{kq}+h-(\epsilon_k-\epsilon_{k-q})/2$,
$E_{kq}=\sqrt{\xi_{kq}^{2}+\Delta^{2}}$,
$\xi_{kq}=(\xi_k+\xi_{k-q})/2$. 
%


The regularized gap equation is
\begin{equation}
-\frac{m}{4\pi
a}=\sum_{\mathbf{k}}\left[\frac{1-f(E_{1\uparrow})-f(E_{1\downarrow})}{2E_{kq}}
\right] \,.
\label{eq:22}
\end{equation}
%
%
Finally, the mean field number equations for $n_{\sigma}$,
are given by
\begin{eqnarray}
n_{\sigma}&=&\sum_{\mathbf{k}}[f(E_{1,\sigma})u_{kq}^{2}
+f(-E_{1,\bar{\sigma}})v_{kq}^{2}] \,.
\label{eq:23}
\end{eqnarray}
Here the coherence factors are
$u_{kq}^{2}=\frac{1}{2}\Big(1+\frac{\xi_{kq}}{E_{kq}} \Big)$ and
$v_{kq}^{2}=\frac{1}{2}\Big(1-\frac{\xi_{kq}}{E_{kq}} \Big)$.

There must be another equation which governs $\mathbf{q}$. This can
be obtained by minimizing the thermodynamical potential or free energy
with respect to $\mathbf{q}$. Equivalently, we will 
derive $\mathbf{q}$ from the $T$-matrix
method described below.

\subsection{$T$-matrix Approach Below $T_c$ : 
One plane wave LOFF state}
\label{sec:2d}

We now use the same $T$-matrix based approach (as we did for the Sarma
case, and as indicated by Table I) to make contact with the one plane
wave LOFF state.  
We take the self energy of the form
\begin{equation}
\Sigma_{\sigma}(K)=-\Delta^{2}G_{0\bar{\sigma}}(-K)
=\frac{\Delta^{2}}{i\omega+\xi_{\mathbf{k-q},\bar{\sigma}}} \,,\nonumber
\label{eq:24}
\end{equation}
so that
\begin{eqnarray}
G^{-1}_{\sigma}(K)&=& i\omega-\xi_{\mathbf{k},\sigma}
-\frac{\Delta^{2}}{i\omega+\xi_{\mathbf{k-q},\bar{\sigma}}}\,.\nonumber
\label{eq:25}
\end{eqnarray}
%
%
%
Then we have

\begin{eqnarray}
G_{\uparrow}(K)&=&\frac{u_{k}^{2}}{i\omega_{n}-E_{1,\uparrow}}+\frac{v_{k}^{2}}{i\omega_{n}+E_{1,\downarrow}} \nonumber \\
G_{\downarrow}(K)&=&\frac{u_{k}^{2}}{i\omega_{n}-E_{2,\downarrow}}+\frac{v_{k}^{2}}{i\omega_{n}+E_{2,\uparrow}}
\label{eq:24a}
\end{eqnarray}
Here $E_{2,\uparrow}=E_{kq}-h-(\epsilon_k-\epsilon_{k-q})/2$ and
$E_{2,\downarrow}=E_{kq}+h+(\epsilon_k-\epsilon_{k-q})/2$. Note if
$\mathbf{k}\to\mathbf{-k+q}$, then $E_{2,\uparrow}\to E_{1,\uparrow}$
and $E_{2,\downarrow}\to E_{1,\downarrow}$.  The number equations are
$n_{\sigma}=\sum_{K}G_{\sigma}(K)$, which yields the same answer
(Eq.~\ref{eq:23}) we found in the mean field approach.
For $T\leq T_{c}$, using the BEC condition
$U^{-1}+\chi(0,\mathbf{q})=0$ we thus arrive at the gap equation
we found earlier in Eq.~(\ref{eq:21}).

We next investigate the pair susceptibility and its extremal value at $
Q= (0,\mathbf{q})$.  We define the contributions to the pair
susceptibility in the presence of population imbalance as in
Eqs.(\ref{eq:1})-(\ref{eq:7}).
For the one plane wave LOFF state, the BEC condition is that
$t(0,\mathbf{p})$ diverges at a non-zero momentum
$\mathbf{p}=\mathbf{q}$. Thus, at this momentum, the quantity
$\chi(0,\mathbf{p})$ should reach a \textit{maximum} at
$\mathbf{p}=\mathbf{q}$. We determine $\mathbf{q}$ by requiring
\begin{equation}
\left.\frac{\partial\chi(0,\mathbf{p})} {\partial\mathbf{p}}
\right|_{\mathbf{p=q}}=0 \,,
\label{eq:26}
\end{equation}
Where $\chi(P)$ is explicitly shown in Eq.~(\ref{eq:LOFF1chi}).
%
%
%
%
This yields
\begin{eqnarray}
\lefteqn{0=\frac{\partial\chi(0,\mathbf{p})}{\partial\mathbf{p}}\Big|_{\mathbf{p=q}}}\nonumber\\
&=&\frac{1}{\Delta^2}\sum_{\mathbf{k}}\bigg\{\frac{\mathbf{q}}{2}\Big[
\Big(1-\frac{\xi_{kq}}{E_{kq}}\Big)-[f(E_{1,\uparrow})+f(E_{1,\downarrow})]\frac{\xi_{kq}}{E_{kq}}\Big] \nonumber\\
&&{}+\left(\mathbf{k-}\frac{\mathbf{q}}{2}\right)[f(E_{1,\uparrow})-f(E_{1,\downarrow})]\bigg\}\,.
\label{eq:27}
\end{eqnarray}
%

It is important to stress, as we show in Appendix \ref{App:LOFF1}, that
this extremal condition on $\chi(P)$ is equivalent to the condition that
the net current is identically zero in this situation.
\begin{eqnarray}
\mathbf{j}&=&\sum_{K}\mathbf{k}[G_{\uparrow}(K)+G_{\downarrow}(K)]=0
\label{eq:28}
\end{eqnarray}
This key
observation shows that the present way of computing $\mathbf{q}$
directly from the pair susceptibility is consistent with the
counterparts in the literature, which are based on the vanishing of the
net current in equilibrium.

At the mean field level we, then, have four unknowns $\Delta$,
$\mathbf{q}$, $\mu$ and $h$, and four equations: two number equations,
the gap equation, and the condition on the vanishing of the first order
derivative.


\section{Mean Field Theory of Two Plane Wave LOFF State}
\label{sec:III}

The two plane wave LOFF state has not been studied in much detail away
from $T_c$ and the associated tri-critical point \cite{LOFF_Review}.
Here we address this state at more general temperatures, starting with
the very natural mean field Hamiltonian in the same spirit as in the
previous two cases.  The Hamiltonian is given by
\begin{eqnarray}
H^{MF}&=&\sum_{\mathbf{k}}\Big\{\xi_{\mathbf{k}\uparrow}c^{\dagger}_{\mathbf{k}\uparrow}c_{\mathbf{k}\uparrow}+\xi_{\mathbf{k}\downarrow}c^{\dagger}_{\mathbf{k}\downarrow}c_{\mathbf{k}\downarrow}\nonumber \\
& &{}+\frac{1}{\sqrt{2}}\Delta(c^{\dagger}_{\mathbf{k}\uparrow}c^{\dagger}_{\mathbf{-k+q}\downarrow}+c_{\mathbf{k}\downarrow}c_{\mathbf{-k+q}\uparrow})\nonumber \\
& &{}+\frac{1}{\sqrt{2}}\Delta(c^{\dagger}_{\mathbf{k}\uparrow}c^{\dagger}_{\mathbf{-k-q}\downarrow}+c_{\mathbf{k}\downarrow}c_{\mathbf{-k-q}\uparrow})\Big\}
\label{eq:29}
\end{eqnarray}

To solve for the value of the gap we generalize the Bogoliubov
diagonalization procedure.
If one rewrites the effective Hamiltonian in the basis set
$\mathbf{B}_{\uparrow}^{T}=(c_{\mathbf{k}\uparrow},
c^{\dagger}_{\mathbf{-k-q}\downarrow},
c^{\dagger}_{\mathbf{-k+q}\downarrow})$ and $\mathbf{B}_{\downarrow}^{T}
=(c_{\mathbf{k}\downarrow}, c^{\dagger}_{\mathbf{-k-q}\uparrow},
c^{\dagger}_{\mathbf{-k+q}\uparrow})$, the Hamiltonian can be written in
two $3\times 3$ matrices.  Here we show the matrix for
$\mathbf{B}_{\uparrow}$:
\begin{equation}
\mathbf{H}_{\uparrow}=\left(\begin{array}{ccc} \xi_{\mathbf{k}\uparrow}
& \frac{1}{\sqrt{2}}\Delta & \frac{1}{\sqrt{2}}\Delta \\ \frac{1}{\sqrt{2}}\Delta & -\xi_{\mathbf{k+q}\downarrow} & 0 \\
\frac{1}{\sqrt{2}}\Delta & 0 & -\xi_{\mathbf{k-q}\downarrow}
\label{eq:30}
\end{array}\right).
\end{equation}

We can equivalently write the $\uparrow$ component of the $3\times 3$
matrix in terms of the basis
$\mathbf{B'}_{\uparrow}^{T}=(c_{\mathbf{k+q}\uparrow},
c_{\mathbf{k-q}\uparrow}, c^{\dagger}_{\mathbf{-k}\downarrow})$
The mean field gap equation is derived from the averages obtained from a
symmetric combination of the two basis sets as
\begin{subequations}
\label{eq:LOFF2_gap}
\begin{equation}
\Delta=\sum_{K} U \langle
c_{k,\uparrow}c_{-k-q,\downarrow}\rangle
=\sum_{K} U \langle
c_{k-q,\uparrow}c_{-k,\downarrow}\rangle
,
\label{eq:31}
\end{equation}
which is equal to
\begin{equation}
\Delta=\sum_{K} U \langle
c_{k,\uparrow}c_{-k+q,\downarrow}\rangle
=\sum_{K} U\langle
c_{k+q,\uparrow}c_{-k,\downarrow}\rangle .
\label{eq:32}
\end{equation}
\end{subequations}
We choose this symmetrized form to write down the gap equation, which
reflects the underlying symmetry of the Hamiltonian.

The resulting gap equation can be shown to be of the form
\begin{eqnarray}
\frac{1}{U}&=&\frac{1}{2}\sum_{\mathbf{k}}\left\{[f(E_{1\uparrow})
+f(E_{1\downarrow})]\frac{E_{1}+\mathcal{E}_{kq}}{(E_{1}-E_{2})
(E_{1}-E_{3})}\right.\nonumber\\
&&{}+[f(E_{2\uparrow})+f(E_{2\downarrow})]\frac{E_{2}
+\mathcal{E}_{kq}}{(E_{2}-E_{1})(E_{2}-E_{3})}\nonumber \\
& &{}+\left.
     [f(E_{3\uparrow})+f(E_{3\downarrow})]\frac{E_{3}+\mathcal{E}_{kq}} 
{(E_{3}-E_{1})(E_{3}-E_{2})}\right\}\,.
\label{eq:33}
\end{eqnarray}
%
Here we define $E_{1}$, $E_{2}$, and $E_{3}$
as the solutions to the cubic equation which is related to
the determinant of the 3x3 eigenvalue equation 
\begin{equation}
x^{3}+\mathcal{E}_{kq}x^{2}-\Big[\tilde{E}_{kq}^{2}+\Big(\frac{\mathbf{k\cdot
q}}{m}\Big)^{2}\Big]x-\mathcal{E}_{kq}\Big[\tilde{E}_ {kq}^{2}-\Big(\frac{\mathbf{k\cdot
q}}{m}\Big)^{2}\Big]=0.
\label{eq:34}
\end{equation}
where 
$\mathcal{E}_{kq}=(k^{2}+\frac{1}{2}q^{2})/2m-\mu$ and $\tilde{E}_{kq}=\sqrt{\mathcal{
E}_{kq}^{2}+\Delta^{2}}$. Then $E_{j\uparrow}=E_{j}-(q^{2}/4m)-h$ and $E_{j\downarrow}=E_{j}-(q^{2}/4m)+h$ for $j=1, 2, 3$.

The mean field number equations are
\begin{eqnarray}
n_{\sigma}&=&\sum_{\mathbf{k}}\Big[\eta_{kq}^{2}
f(E_{1\sigma})+\gamma_{kq}
^{2}f(E_{2\sigma})\nonumber \\
&&{}+(1-\eta_{kq}^{2}-\gamma_{kq}^{2})f(E_{3\sigma})
\Big]
\label{eq:35}
\end{eqnarray}
where we define 
\begin{eqnarray}
\eta_{kq}^{2}&=&\frac{(E_{1}+\mathcal{E}_{kq})^{2}-\big(\frac{\mathbf{k\cdot
    q}}{m}\big)^{2}}{(E_{2} -E_{1})(E_{3}-E_{1})} \\ \nonumber
\gamma_{kq}^{2}&=&\frac{(E_{2}+\mathcal{E}_{kq})^{2}-\big(\frac{\mathbf{k\cdot
    q}}{m}\big)^{2}} {(E_{1}-E_{2})(E_{3}-E_{2})}
\label{eq:36}
\end{eqnarray}

\subsection {Green's Function Approach Below $T_c$: Two plane wave LOFF state}
\label{sec:IIIA}

For the counterpart mean field theory, approached from a $T$-matrix
scheme, we define the self energy as
\begin{eqnarray}
\Sigma_{\sigma}(K)&=&-\frac{1}{2}\Big[\Delta^{2}G_{0\bar{\sigma}}
(-K+Q)+\Delta^{2}
G_{0\bar{\sigma}}(-K-Q) \Big] \nonumber \\
&=&\frac{1}{2}\Big[\frac{\Delta^{2}}{i\omega_{n}+\xi_{\mathbf{k-q}\bar{\sigma}}}
+\frac{\Delta^{2}}{i\omega_{n}+\xi_{\mathbf{k+q}\bar{\sigma}}} \Big]
\label{eq:37}
\end{eqnarray}
Here $Q=(0,\mathbf{q})$.


The Green's function is then given 
by
\begin{eqnarray}
G_{\sigma}(K)&=&\frac{1}{G_{0\sigma}(K)-\Sigma_{\sigma}(K)} \\
&=&\frac{(i\omega_{n}+\xi_{\mathbf{k+q}\bar{\sigma}})(i\omega_{n}+\xi_{\mathbf{k-q
}\bar{\sigma}})}{(i\omega_{n}-E_{1\sigma})(i\omega_{n}-E_{2\sigma})(i\omega_{n
}-E_{3\sigma})} \nonumber \\
&=&\frac{\eta_{kq}^{2}}{i\omega_{n}-E_{1\sigma}}+\frac{\gamma_{kq
}^{2}}{i\omega_{n}-E_{2\sigma}}+\frac{1-\eta_{kq}^{2}
-\gamma_{kq}
^{2}}{i\omega_{n}-E_{3\sigma}} \nonumber
\label{eq:38}
\end{eqnarray}
Again, following Eqs. (\ref{eq:1})-(\ref{eq:7}) we can write out the
form of the pair susceptibility. For the two plane wave LOFF
state, this equation is presented as Eq.~(\ref{eq:LOFF2symchi}) in
Appendix \ref{App:LOFF2}. This quantity, in turn, enters the gap
equation,
%
given by the BEC condition, $1+U\chi(0,\mathbf{q})=0$.  We may write
this gap equation in compact form as
\begin{equation}
\frac{1}{U}=\frac{1}{2}\sum_{K,\sigma}\frac{i\omega_{n}
+\xi_{\mathbf{kq}\bar{\sigma}}}{
(i\omega_{n}-E_{1\sigma})(i\omega_{n}-E_{2\sigma})(i\omega_{n}
-E_{3\sigma})} 
\label{eq:39}
\end{equation}
This will, in turn, reduce to the gap equation (\ref{eq:33}) we deduced
directly from the $3\times 3$ matrix analysis.

Similarly, in our Green's function formalism, we have
$n_{\sigma}=\sum_{K}G_{\sigma}(K)$, which reduces to the number
equations we found in Eq.~(\ref{eq:35}).

Next we determine the momentum $\mathbf{p}$ that maximizes
$\chi(P)$ when $\Omega=0$, i.e., we need to find a solution to
$\frac{\partial \chi(0,\mathbf{p})}{\partial\mathbf{p}}=0$.  Since
the Green's function is symmetric under $\mathbf{k}\to -\mathbf{k}$, it
can be shown that $\mathbf{p}=0$ is a solution, which corresponds to the
Sarma phase.  Here, however, we are interested in a LOFF-like state
$\mathbf{p}=\pm\mathbf{q}$.  where both signs contribute in a symmetric
fashion.  Thus we choose $\mathbf{p}=\mathbf{q}$, and, thereby, arrive
at the defining equation for the net momentum of the pairs.

\begin{widetext}
\begin{eqnarray}
0&=&\sum_{K,\sigma}\Big(\frac{\mathbf{k-q}}{m}\Big)\Big\{\frac{i\omega_{n}+\xi_{\mathbf{k+q}\bar{\sigma}}}{i\omega_{n}+\xi_{\mathbf{k-q}\bar{\sigma}}}\,\frac{1}{(i\omega_{n}-E_{1\sigma})(i\omega_{n}-E_{2\sigma})(i\omega_{n}-E_{3\sigma})}\Big\}
\nonumber \\
&=&\sum_{\mathbf{k}}\Big(\frac{\mathbf{k-q}}{m}\Big)\Big\{\frac{2\big(\frac{\mathbf{k\cdot
q}}{m}\big)[f(\xi_{\mathbf{k-q}\uparrow})+f(\xi_{\mathbf{k-q}\downarrow})-2]}{(E_{1}+\mathcal{E}_{kq}-\frac{\mathbf{k\cdot
q}}{m})(E_{2}+\mathcal{E}_{kq}-\frac{\mathbf{k\cdot
q}}{m})(E_{3}+\mathcal{E}_{kq}-\frac{\mathbf{k\cdot q}}{m})} \nonumber \\ 
& &{}+\frac{f(E_{1\uparrow})+f(E_{1\downarrow})}{(E_{1}-E_{2})(E_{1}-E_{3})}\,\frac{E_{1}+\mathcal{E}_{kq}+\frac{\mathbf{k\cdot
q}}{m}}{E_{1}+\mathcal{E}_{kq}-\frac{\mathbf{k\cdot
q}}{m}}+\frac{f(E_{2\uparrow})+f(E_{2\downarrow})}{(E_{2}-E_{1})(E_{2}-E_{3})}\,\frac{E_{2}+\mathcal{E}_{kq}+\frac{\mathbf{k\cdot
q}}{m}}{E_{2}+\mathcal{E}_{kq}-\frac{\mathbf{k\cdot q}}{m}} \nonumber \\ 
& &{}+\frac{f(E_{3\uparrow})+f(E_{3\downarrow})}{(E_{3}-E_{1})(E_{3}-E_{1})}\,\frac{E_{3}+\mathcal{E}_{kq}+\frac{\mathbf{k\cdot
q}}{m}}{E_{3}+\mathcal{E}_{kq}-\frac{\mathbf{k\cdot q}}{m}}\Big\} \,.
\label{eq:40}
\end{eqnarray}
\end{widetext}
This determines the magnitude $\mathbf{q}$; for definiteness we take the
direction of $\mathbf{q}$ as along the $z$-axis.

The case of the single plane wave LOFF state should be contrasted.
There we showed that there was an intimate relation between the
condition that there be no net equilibrium current and the extremal
requirement on $\chi(P)$.  For the two plane wave LOFF state the current
can be shown to be identically zero.  Rather, the only condition one has
to determine $\mathbf{q}$ is the vanishing of the first derivative of
the pair susceptibility.  A similar condition was imposed in the
original Larkin-Ovchinnikov (LO) paper \cite{FFLO}.

\begin{figure}[b]
\begin{center}
\includegraphics[width=0.45\textwidth]{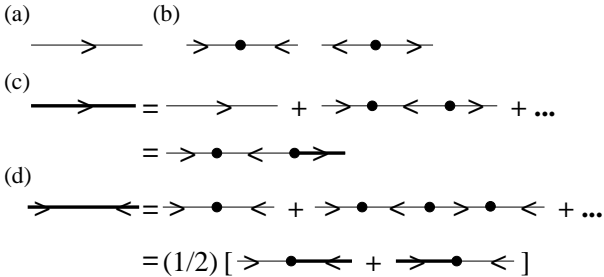}
\end{center}
\caption{\label{fig:BCSdia}Diagrams for Sarma states or alternatively
one plane wave LOFF states including (a) Bare Green's function
$G_{0}$. (b) Vertices from pairing.  The momenta are
$(\mathbf{k}\downarrow, \mathbf{-k}\uparrow)$ and $(\mathbf{k}\uparrow,
\mathbf{-k}\downarrow)$, while for single plane-wave LOFF states the
momenta are $(\mathbf{k}\downarrow, \mathbf{-k+q}\uparrow)$ and
$(\mathbf{k}\uparrow, \mathbf{-k+q}\downarrow)$.  (c) Green's function
$G$.  (d) $\langle c_{\mathbf{k}}\uparrow
c_{\mathbf{-k}\downarrow}\rangle$ for Sarma states and $\langle
c_{\mathbf{k}}\uparrow c_{\mathbf{-k+q}\downarrow}\rangle$ for single
plane-wave LOFF states. Here the thin and thick lines represent bare and full Green's functions, respectively.}
\end{figure}


\begin{figure}[b]
\begin{center}
\includegraphics[width=0.45\textwidth]{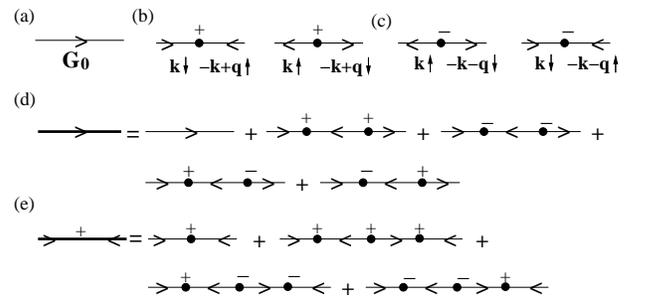}
\end{center}
\caption{\label{fig:LOdia}Diagrams representing the original
 Ovchinnikov-Larkin theory for (a) Non-interacting
Green's function $G_{0}$. (b) Vertices from pairing with $\mathbf{q}$.
(c) Vertices from pairing with $-\mathbf{q}$. (d) Green's function $G$.
(e) $\langle c_{\mathbf{k}}\uparrow c_{\mathbf{-k+q}\downarrow}\rangle$.}
\end{figure}

\subsection {Diagrammatic Interpretation of Green's Function
Approaches: Comparison with Larkin-Ovchinnikov}
\label{sec:IIIB}

We now want to compare the present approach for the two plane wave LOFF
state with that proposed in the original LO paper
\cite{FFLO}. We will do this comparison within a diagrammatic framework.
It is useful first to illustrate the diagrammatic scheme by referring to
the simpler Sarma and one plane wave LOFF states.  These two states have
a rather similar diagrammatic formulation.  We write these diagrams in a
consolidated form in Fig.~\ref{fig:BCSdia}.  The first line indicates
the bare quantities, that is, the bare Green's function $G_0$ in part
(a) and the quantity $G_{0,\uparrow}(K) \Delta G_{0,\downarrow}(-K)$ in
part (b).  This is written for the Sarma state and readily generalized
to the one plane wave LOFF phase.

The second line indicates the (diagonal) dressed Green's function, while
the third line shows effectively the Gor'kov $F$ function.  It should be
noted for Fig.~\ref{fig:BCSdia}d that this latter ``anomalous" or
off-diagonal Green's function is explicitly seen to depend on a
symmetrized sum of the product $G$ and $G_0$.  It is this combination
which we have seen appear in the pair susceptibility $\chi(P)$.  Indeed,
throughout this paper we have found that $GG_0$ is essential for
arriving at the standard mean field theoretic approach.  We can, thus,
conclude from the last line in the figure that the $F$ function is given
by a spin symmetrized combination of $\Delta G G_0$.

\begin{figure}[t]
\begin{center}
\includegraphics[width=0.45\textwidth]{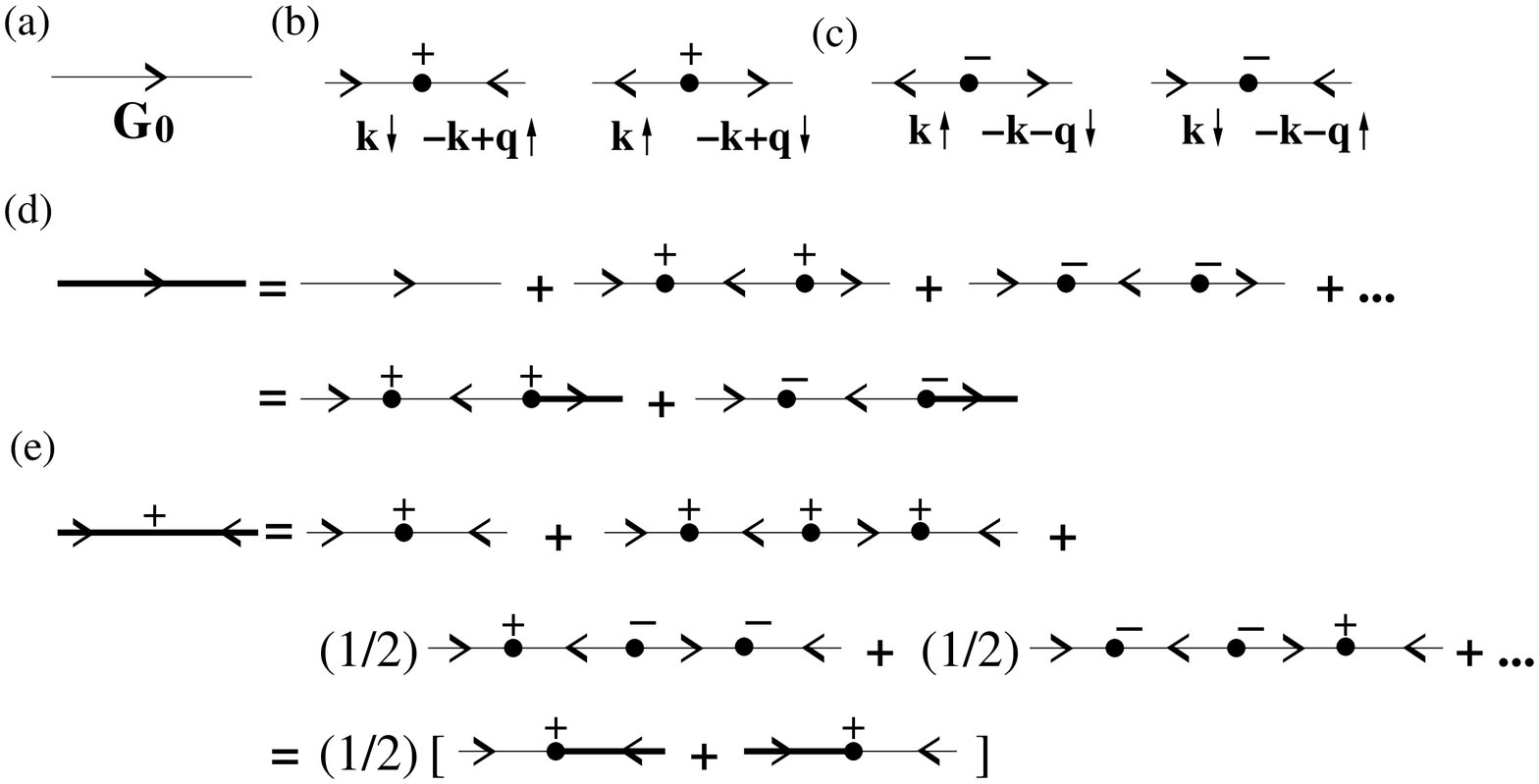}
\end{center}
\caption{\label{fig:QCdia}Diagrams associated with the
3x3 Bogoliubov diagonalization theory for (a) Non-interacting Green's 
function $G_{0}$. (b) Vertices from pairing with $\mathbf{q}$. 
(c) Vertices from pairing with $-\mathbf{q}$. (d) Green's function $G$. 
(e) $\langle c_{\mathbf{k}}\uparrow c_{\mathbf{-k+q}\downarrow}\rangle$.}
\end{figure}

In Fig.~\ref{fig:LOdia} we show the two plane wave LOFF diagrams
originally proposed by LO.  In order to discuss their implications, we
present the diagrams associated with the mean field matrix scheme of the
previous section which are plotted in Fig.\ref{fig:QCdia}.  As in Figure
\ref{fig:BCSdia}, one can see from the last line in part e of
Fig.\ref{fig:QCdia} that the off-diagonal Green's function depends on a
symmetrized sum of $G$ and $G_0$.
One can also see that there are similarities as well as differences in
these latter two approaches.  In the approach of Section \ref{sec:IIIA},
just as in LO we restrict our calculations to diagrams that do not
contain propagators with momentum beyond $\mathbf{\pm k}$ and
$\mathbf{\pm k\pm q}$.  The differences are also apparent.  In the LO
paper, the (diagonal component of the) Green's function diagrams were
summed up to second order in $\Delta$. In the gap equation LO dropped
diagrams containing propagators with higher momenta and truncated the
series at third order in $\Delta$.

If one were to follow the original LO scheme, but sum the entire series,
then we will arrive at a slightly modified gap equation.  We would find
instead the gap is determined by
%
\begin{widetext}
\begin{eqnarray}
\frac{1}{U}&=&\sum_{K}\Big\{[G_{0\uparrow}(-K+Q)G_{\downarrow}(K)
+G_{0\downarrow}(-K+Q)G_{\uparrow}(K)]
-\frac{1}{2}[G_{0\uparrow}(-K+Q)G^{(1)}_{\downarrow}(K)+G_{0\downarrow}(-K+Q)
G^{(1)}_{\uparrow}(K)]\Big\} \nonumber \\
&=&\sum_{\mathbf{k}}\Big\{[f(E_{1\uparrow})
+f(E_{1\downarrow})]\frac{E_{1}+\mathcal{E}_{kq}}{(E_{1}-E_{2})
(E_{1}-E_{3})}+[f(E_{2\uparrow})+f(E_{2\downarrow})]\frac{E_{2}
+\mathcal{E}_{kq}}{(E_{2}-E_{1})(E_{2}-E_{3})} \nonumber \\
& &{} +[f(E_{3\uparrow})+f(E_{3\downarrow})]\frac{E_{3}+\mathcal{E}_{kq}}
{(E_{3}-E_{1})(E_{3}-E_{2})}\Big\}-\sum_{\mathbf{k}}
\frac{f(E^{(1)}_{1,\uparrow})+f(E^{(1)}_{1,\downarrow})-1}{2E^{(1)}_{kq}}
\label{eq:41}
\end{eqnarray}
\end{widetext}
and thus the pair susceptibility derived from this method is
\begin{eqnarray}
\lefteqn{\chi(P)}\nonumber\\
&&=\frac{1}{2}\sum_{K}\Big\{\Big[G_{0\uparrow}(P-K)G_{\downarrow}(K)+G_{0\downarrow}(P-K)G_{\uparrow}(K)\Big]\nonumber\\
&&{}-\Big[G_{0\uparrow}(P-K)G^{(1)}_{\downarrow}(K)+G_{0\downarrow}(P-K)G^{(1)}_{\uparrow}(K)\Big]\Big\}.\nonumber\\
\label{eq:42}
\end{eqnarray}
We could similarly use this modified pair susceptibility to arrive at
the constraint on the value of $\mathbf{q}$.
We discuss these contributions in Appendix \ref{App:LOFF2}.
For the above equation we define
the Green's function for the single plane-wave LOFF states
as $G^{(1)}_{\uparrow}(K)=(i\omega_{n}+\xi_{\mathbf{k-q}\downarrow})
/(i\omega_{n}-E^{(1)}_{1,\uparrow})(i\omega_{n}+E^{(1)}_{1,\downarrow})$
and 
$G^{(1)}_{\downarrow}(K)=(i\omega_{n}+\xi_{\mathbf{k-q}\uparrow})
/(i\omega_{n}-E^{(1)}_{2,\downarrow})(i\omega_{n}+E^{(1)}_{2,\uparrow})$.
The energy spectrum of the single plane-wave LOFF states is
$E^{(1)}_{1,\uparrow}=E^{(1)}_{\mathbf{kq}}+
\frac{1}{2m}\Big(\mathbf{k\cdot q}-\frac{q^{2}}{2}\Big)-h$, 
$E^{(1)}_{1,\downarrow}=E^{(1)}_{\mathbf{kq}}-\frac{1}{2m}\Big(\mathbf{k\cdot q}-
\frac{q^{2}}{2}\Big)+h$, $E^{(1)}_{2,\uparrow}=E^{(1)}_{\mathbf{kq}}-
\frac{1}{2m}\Big(\mathbf{k\cdot q}-\frac{q^{2}}{2}\Big)-h$, 
and $E^{(1)}_{2,\downarrow}=E^{(1)}_{\mathbf{kq}}+
\frac{1}{2m}\Big(\mathbf{k\cdot q}-\frac{q^{2}}{2}\Big)+h$, 
where $E^{(1)}_{\mathbf{kq}}=\sqrt{(\mathcal{E}_{kq}-\frac{\mathbf{k\cdot q}}{2m})^{2}
+\Delta_{1}^{2}}$ and we define $\Delta_{1}^{2}=\frac{1}{2}\Delta^{2}$.

In summary, to make contact with the original results of LO, one must
subtract a second symmetrized term, which represents pairing with only
one momentum. It would thus appear, that relative to LO, there is an
overcounting in the summation of the two series from the diagrams shown
in Fig.\ref{fig:QCdia}, and associated with the 3x3 matrix or mean field
representation.
At this stage it is difficult to determine which of the two
plane wave LOFF representations is the more appropriate. It will
be essential in future to study them both numerically in the presence of BCS-BEC
crossover effects.

\section{Beyond Simple Mean Field Theory: Pair Fluctuation Effects}
\label{sec:IV}

\subsection{Inclusion of the Pseudogap}

For definiteness the equations which appear in this section apply to the
Sarma-like phase. We can readily generalize to include the two different
LOFF states.

\begin{figure}
\centerline{\includegraphics[clip,width=3.0in]{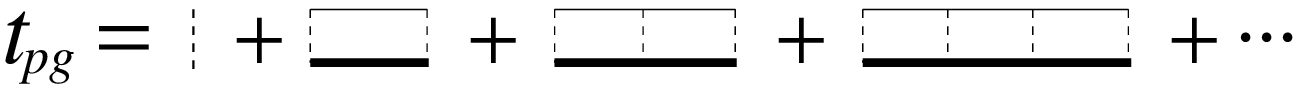}}
\caption{$T$-matrix diagrams. Note that there is one dressed and
full Green's function. This represents the propagator for the 
non-condensed pairs.}
\label{fig:4}
\end{figure}

\begin{figure}
\centerline{\includegraphics[clip,width=3.in]
{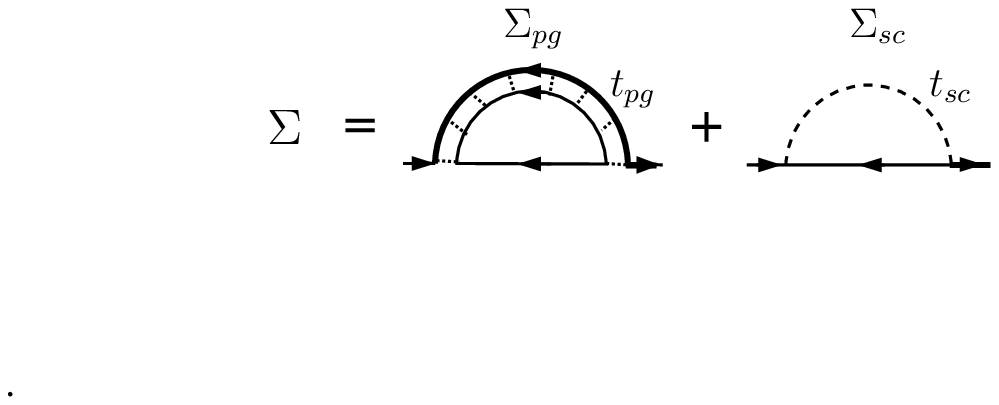}}
\caption{Self energy diagram for the present $T$-matrix scheme,
indicating the condensed ($\Sigma_{sc}$) and non-condensed
($\Sigma_{pg}$) pairs.}
\label{fig:5}
\end{figure}

This diagrammatic representation of our $T$-matrix scheme is shown in
Figs.~\ref{fig:4} and \ref{fig:5}.  The first of these indicates the
propagator for non-condensed pairs which we refer to as $t_{pg}$, and
the second of these the total self energy.  One can see throughout the
combination of one dressed and one bare Green's function,
as represented by the thick and thin lines.
%
The self energy consists of two contributions from the non-condensed
pairs or pseudogap ($pg$) and from the condensate ($sc$).  There are,
analogously, two contributions in the full $T$-matrix
\begin{eqnarray}
t &=& t_{pg} + t_{sc} \,, \label{t-matrix}\\
t_{pg}(P)&=& \frac{U}{1+U \chi(P)}, \qquad P \neq 0 \,,
\label{t-matrix_pg}\\
t_{sc}(P)&=& -\frac{\Delta_{sc}^2}{T} \delta(P) \,,
\label{t-matrix_sc}
\label{eq:43}
\end{eqnarray}
where we write
$\Delta_
{sc}=-U \sum _{\bf k}\langle c_{-{\bf k}\downarrow}c_{{\bf
k}\uparrow}\rangle$.
Similarly, we have for the fermion self energy
\begin{equation}
\Sigma_\sigma (K) = \Sigma_\sigma ^{sc}(K) + \Sigma_\sigma ^ {pg} (K) = \sum_P t(P) G_{0,\bar{\sigma}} (P-K) \,.
\label{eq:sigma2}
\end{equation}
 We can see at once that
\begin{equation}
\Sigma_\sigma ^{sc}(K) = \sum_P t_{sc}(P) G_{0,\bar{\sigma}}(P-K) =
-G_{0,\bar{\sigma}} (-K) \Delta_{sc}^2 
\,.
\label{eq:70}
\end{equation}
The vanishing of the pair chemical potential implies
that
\begin{equation}
t_{pg}^{-1} (0) = U^{-1} + \chi(0) = 0, \qquad T \leq T_c \;.
\label{eq:3e}
\end{equation}
Moreover, a vanishing chemical potential means that $t_{pg}(P)$
is strongly peaked around $P=0$. Thus, we may approximate \cite{Maly1}
Eq. (\ref{eq:sigma2}) to yield
\begin{equation}
\Sigma_{\sigma} (K)\approx -G_{0,\bar{\sigma}} (-K) \Delta^2 \,,
\label{eq:sigma3}
\end{equation}
where
\begin{equation}
\Delta^2 (T) \equiv \Delta_{sc}^2(T)  + \Delta_{pg}^2(T) \,,
\label{eq:sum}
\end{equation}
Importantly, we are led to identify the quantity $\Delta_{pg}$
\begin{equation}
\Delta_{pg}^2 \equiv -\sum_{P\neq 0} t_{pg}(P).
\label{eq:delta_pg}
\end{equation}
Note that in the normal state (where $\mu_{pair}$ is non-zero)
Eq. (\ref{eq:sigma3}) is no longer a good approximation.

We now have a closed set of equations for addressing the ordered phase.
We can similarly extend this approach to temperatures somewhat above
$T_c$, by self consistently including a non-zero pair chemical
potential.  This is a necessary step in addressing a trap as
well \cite{ChienRapid}.  Additionally, the propagator for non-condensed
pairs can now be quantified, using the self consistently determined pair
susceptibility.
At small four-vector $P$, we may expand the inverse of the $T$-matrix,
after analytical continuation ($i\Omega_l \rightarrow \Omega+i0^+$), to
obtain
\begin{equation}
a_1\Omega^2 + Z(\Omega - \frac{p^2}{2 M^*} 
+ \mu _{pair}  +i \Gamma_P),
\label{Omega_q:exp}
\end{equation}
where the imaginary part $\Gamma_P \rightarrow 0$ rapidly as
$p\rightarrow 0$ below $T_c$.  Because we are interested in the moderate
and strong coupling cases, we drop the $a_1 \Omega^2$ term in
Eq. (\ref{Omega_q:exp}), and hence
\begin{equation}
t_{pg}(P) = \frac { Z^{-1}}{\Omega - \Omega_p +\mu_{pair} + i \Gamma_P},
\label{eq:expandt}
\end{equation}
where we associate
\begin{equation}
\Omega_{\mathbf{p}} \equiv \frac{p^2} {2 M^*} .
\label{eq:53}
\end{equation}
This establishes a quadratic dispersion and defines the effective pair
mass, $M^*$.

Finally, one can rewrite Eq. (\ref{eq:delta_pg})
as
\begin{equation}
\Delta_{pg}^2 (T) = Z^{-1}\sum_{\mathbf{p}} b(\Omega_p) \,,
\label{eq:81}
\end{equation}
where $b(x)$ is the Bose distribution function.
Analytical expressions for this mass are possible via a
small ${\bf p}$ expansion of $\chi$.
In this way we find
\begin{eqnarray}
\chi(P)-\chi(0) &\approx& Z \left(\Omega  - \frac{p^2}{2M^*}\right),
\label{eq:55}
\end{eqnarray}
%
The coefficients are
\begin{equation}
Z=\left.\frac{\partial \chi}{\partial\Omega}\right|_{\Omega=0,\mathbf{p}=0}
\label{eq:56}
\end{equation}
and
\begin{equation}
\frac{1}{2M^*} =-\left.\frac{1}{6Z}\frac{\partial^{2} \chi}{\partial
  p^{2}}\right|_{\Omega=0,\mathbf{p}=0} \,.
\end{equation}

\subsection{Pair Dispersion in Sarma State and Physical
Consequences at Finite $T$}
\label{sec:IVA}

The theoretical framework in the previous section can now be implemented
to address the physics at finite $T$. We do this here only for the Sarma
state, thereby expanding on earlier papers \cite{Chien06,ChienRapid}.
The counterpart results for the two LOFF phases are presented in
the two appendices.

We begin by determining the form of the pair dispersion.
The pair susceptibility is
given by
\begin{eqnarray}
\lefteqn{\chi(P)=\frac{1}{2}\big[\chi_{\uparrow\downarrow}(P)
+\chi_{\downarrow\uparrow}(P)\big]} \nonumber \\
&&=\sum_{\mathbf{k}}\bigg\{ u_{k}^{2}\frac{\bar{f}(E_{k})
+\bar{f}(\xi_{p-k})-1}{i\Omega-\xi_{p-k}-E_{k}} 
+v_{k}^{2}\frac{\bar{f}(\xi_{p-k})-
\bar{f}(E_{k})}{i\Omega-\xi_{p-k}+E_{k}}\bigg\}.\nonumber\\
\end{eqnarray}
Inserting this form into the $T$-matrix [Eq.~(\ref{eq:1})]
we readily find that the coefficent $Z$ is
\begin{eqnarray}
Z&=&\frac{\partial \chi}{\partial\Omega}\Big|_{\Omega=0,\mathbf{p}=0} \nonumber \\
&=&\frac{1}{2\Delta^{2}}\left[n-2\sum_{\mathbf{k}}
\bar{f}(\xi_{k})\right].
\end{eqnarray}
Here $n=\sum_{\mathbf{k},\sigma}n_{\sigma}(k)$ is the total density.

Next we want to calculate $\Omega_{p}$ (defined above) and its derivatives.
\begin{eqnarray}
\Omega_{p}&=&-\frac{1}{Z}t^{-1}|_{\Omega=0} \nonumber \\
&=&-\frac{1}{Z}\Bigg\{\frac{m}{4\pi a}-\sum_{\mathbf{k}}
\bigg[\frac{1}{2\epsilon_{k}} 
+u_{k}^{2}\frac{1-\bar{f}(E_{k})-\bar{f}(\xi_{p-k})}
{E_{k}+\xi_{p-k}} \nonumber \\
&&{}+
v_{k}^{2}\frac{\bar{f}(\xi_{p-k})-\bar{f}(E_{k})}{E_{k}-\xi_{p-k}}\bigg]\Bigg\} 
\end{eqnarray}

Since $\xi_{k}=k^{2}/2m-\mu$, we have $\nabla_{k}\xi_{k}=\mathbf{k}/m$
and $\nabla_{k}^{2}\xi_{k}=3/m$.
Then
\begin{eqnarray}
\frac{1}{2M^*}&=&\frac{1}{6}\frac{\partial^{2}\Omega_{p}}{\partial p^{2}}\Big|_{\mathbf{p}=0}
\nonumber \\
&=&-\frac{1}{12mZ\Delta^{2}}\sum_{\mathbf{k}}\Big\{\bar{f}(\xi_{k})\Big[6+\frac{8\xi_{k}(\xi_{k}+\mu)}{m\Delta^{2}}\Big]
\nonumber \\
& &{}-\bar{f}(E_{k\uparrow})\Big[\frac{6\xi_{k}}{E_{k}}+
\frac{4(\xi_{k}+\mu)(E_{k}^{2}+\xi_{k}^{2})}{mE_{k}\Delta^{2}}\Big] \nonumber \\
& &{}+4\bar{f}^{\prime}(\xi_{k})\frac{(\xi_{k}+\mu)}{m}-3\Big(1-\frac{\xi_{k}}{E_{k}}\Big) \nonumber \\
& &{}+\frac{2E_{k}}{m\Delta^{2}}\Big(1-\frac{\xi_{k}}{E_{k}}\Big)^{2}
(\xi_{k}+\mu)\Big\}.
\end{eqnarray}
Here $\bar{f}^{\prime}(x)$ is the derivative of $\bar{f}(x)$.

With this dispersion, then one can compute the number of
non-condensed pairs,
 $n_{pair}=Z\Delta_{pg}^{2}=\sum_{\mathbf{p}} b(\Omega_{\mathbf{p}})$.
We then have
\begin{equation}
\Delta_{pg}^{2}=\frac{(2M^*T)^{3/2}}{2\pi^{2}Z}
\frac{\sqrt{\pi}}{4}\zeta\Big(\frac{3}{2}\Big),
\end{equation}
where $\zeta(x)$ is the Riemann zeta function.
When $T>T_{c}$, $\Delta_{pg}^{2}=Z^{-1}\sum_{\mathbf{p}}b(\Omega_{\mathbf{
p}}-\mu_{pair})$.

Physically, these non-condensed pairs, which appear at finite $T$, have
been shown to have important physical consequences.  In the homogeneous
situation, we have found that finite $T$ stabilizes the superfluid state
leading to an ``intermediate temperature superfluid"
\cite{Chien06,Stability}. 
In addition, the presence of a finite excitation gap at $T_c$
[$\Delta_{pg}(T_c) \neq 0$] makes the behavior of the superfluid
transition temperature more complex than its mean field counterpart,
$T_c^{MF}$. We frequently find a double-valued structure \cite{Chien06},
with superfluidity existing only for temperature intermediate between
the two $T_c$'s.

In the trapped situation, we have found \cite{ChienRapid} that these
pairs enter explicitly as the mechanism for carrying polarization within
the Sarma phase.  At low $T$ there is very little polarization carried
by the superfluid core; instead, the region of the trap where
$\Delta_{sc}=0$, but $\Delta \neq 0$ -- which can be called ``the mixed
normal region"-- carries the bulk of the polarization, much as observed
experimentally \cite{ZSSK206}.

\section{PHYSICAL PICTURE: FERMIONIC PAIRING IN ANALOGY TO BEC}
\label{sec:V}

We now return to the strong analogies between this BCS-based or Leggett
mean field theory \cite{Leggett} and Bose condensation of point bosons,
as summarized in Table I.

We have three central equations.

1. The pair chemical potential must vanish at and below
$T_c$
\begin{equation}
\mu_{pair} = 0, \quad (T \leq T_c).
\label{eq:57}
\end{equation}
Importantly this condition leads to the mean field gap equations derived
in Sections \ref{sec:II} and \ref{sec:III}. These gap equations then
provide a specific value for $\Delta(T)$, according to the different
phases being contemplated.

2. There must be a conservation of the total number of (composite)
``bosons" in the system.  For this condition, our
central equation is Eq.~(\ref{eq:sum}).
Here it is understood that the number of "bosons" is effectively
represented by the parameter $\Delta^2 (T)$. In the fermionic limit,
this parameter reflects the number of bosons through the energy which is
needed to create fermions, and thereby break the bosons apart.  Unlike
the point boson case, here the ``total boson number" is temperature
dependent and has to be self-consistently determined. As expected, in
the deep BEC regime, where the fermionic excitations are negligible, the
pair density is given by $n_{pair} = Z \Delta^2 \approx n/2$.

3. The number of non-condensed pairs is readily computed in terms of the
pair dispersion, just as in conventional BEC. For this condition our
central equation is Eq.~(\ref{eq:81}).

Then, just as in conventional BEC, the number of condensed bosons
(proportional to $\Delta_{sc}^2$) is determined by the difference
between $\Delta^2(T)$ and $\Delta_{pg}^2 (T)$.  This, in turn,
determines the transition temperature $T_c$ as the lowest 
temperature(s) in the normal state at which noncondensed pairs exhaust
the total weight of $\Delta^2$ so that $\Delta_{pg}^2 = \Delta^2$.
Solving for the ``transition temperature" in the absence of pseudogap
effects \cite{Machida2,YD05,LatestStoof} leads to the quantity
$T_c^{MF}$.  More precisely, $T_c^{MF}$ should be thought of as the
temperature at which the excitation gap $\Delta(T)$ vanishes.  This
provides a reasonable estimate, for the pairing onset temperature $T^*$,
(when a stable superfluid phase exists).  This is distinguished
from the transition temperature. We note that $T^*$
represents a smooth crossover rather than an abrupt phase transition.

It should be stressed that the dispersion relation for the non-condensed
pairs is quadratic. This appears in the Sarma and both one- and
two-plane wave LOFF states.  For the latter two states, it can be seen
to be closely related to the fact that $\chi(0, \mathbf{q})$ must
reach a maximum at $\mathbf{q}$. While
\textit{one will always find a linear dispersion in the collective mode
spectrum \cite{Kosztin2}}, within the present class of BCS-BEC crossover
theories, the restriction to a $T$-matrix scheme means that there is no
feedback from the collective modes onto the pair excitation spectrum.
In effect, the $T$-matrix approximation does not incorporate pair-pair
interactions at a level needed to arrive at this expected linear
dispersion in the pair excitation spectrum.  Nevertheless, because
essentially all theories which address population imbalance build on the
simplest BCS-Leggett mean field theory, there is good reason to first
address this level of approximation when including finite temperature
effects.

\section{Summary}
\label{sec:VI}

It should be clear from the previous sections that the zero and finite
temperature theories of population imbalanced superfluids can be
consolidated into one general theory, based on the quantity which we
call the pair susceptibility, $\chi(P)$. This has a very specific form
in the class of mean field theories currently applied to address
population imbalance.  The quantity $\chi(P)$ enters into the propagator
for non-condensed pairs Eq.~(\ref{t-matrix_pg}) which is just the
$T$-matrix, $t(P)$.  When non-condensed pairs are in equilibrium with a
condensate, they must have zero chemical potential.  This, in turn,
yields the various gap equations for the Sarma and the one- and
two-plane-wave LOFF states, provided one take a special form for the
pair susceptibility involving one dressed and one bare Green's function.

Importantly, the same pair propagator characterizes the effective number
of non-condensed pairs as seen in Eq.~(\ref{eq:delta_pg}).  Each of these
population imbalanced superfluids has an associated $T \neq 0$ pseudogap
contribution, which is summarized for the Sarma case in Section
\ref{sec:III} and for the two LOFF cases in Appendices \ref{App:LOFF1}
and \ref{App:LOFF2}.  This pseudogap contribution serves to
differentiate the order parameter from the gap parameter at all non-zero
$T$, and in all cases except strict BCS theory.  At a more physical
level, in the normal state there is an excitation gap for fermionic
excitations associated with ``pre-formed" pairs.  In the superfluid
phase, there is a new form of condensate excitation, not found in BCS
theory and associated with excited pair states. As the pairing
attraction becomes stronger, it pays to excite pairs of atoms rather
than create single fermion excitations which cost an energy gap.  These
pseudogap effects are an essential component of BCS-BEC crossover
theory, and they are necessary in order to smoothly involve from the
fermionic statistics of BCS to the bosonic statistics of BEC.
Interestingly, they are also widely observed in high temperature
superconductors \cite{ourreview,ReviewJLTP}.

Thus far, we have applied the present theoretical formalism to the Sarma
state both in the homogeneous \cite{Chien06} and trapped
\cite{ChienRapid} configurations. From the point of view of comparing
with experiment, our trapped calculations have again underlined the
importance of pseudogap effects.  We find that the bulk of the
polarization is carried in the pseudogap region of the trap: outside the
condensate but in the region where both spin states (and thus pairing)
are present.  Because we find that only modest polarizations are stable,
it appears necessary in future to include LOFF-like condensates as well,
although their stability must be demonstrated \cite{Stability}.  This
observation is also consistent with numerical calculations
\cite{Machida2,Kinnunen} based on Bogoliubov-de Gennes theory.  A key
contribution of the present work is that it lays the groundwork for
addressing the one and two plane wave LOFF phases, at general
temperature.

In summary, this paper has presented a theoretical formalism for
the Sarma and one- and two-plane wave LOFF states at zero and finite $T$
as one varies from BCS to BEC, in the presence of an arbitrary
population imbalance.  Our premise is that the effects of finite $T$,
which necessarily must be accomodated in any comparison with experiment,
must be compatible with the $T=0$ formalism.  The zero temperature
formalism we use here reduces to the standard one in the literature
\cite{SR06,PWY05} for the Sarma and one-plane wave LOFF states.
However, our studies of the two-plane wave LOFF state present new
results by extending the current literature away from the
Landau-Ginzburg regime (near $T_c$).

\acknowledgments

This work was supported by NSF PHY-0555325
and NSF-MRSEC Grant No.~DMR-0213745. 


\appendix

\section{Additional Results for the One Plane Wave LOFF State}
\label{App:LOFF1}

The following two appendices are dedicated to presenting additional
details on the one- and two-plane-wave LOFF states. The results are
presented in appendix form both for clarity and to avoid some of the
more technical details in the main text.  First, for the one-plane-wave
LOFF state, we wish to explore the relationship between the zero current
condition applied by Fulde and Ferrell \cite{FFLO} and the extremal
condition of $\chi(0,\mathbf{q})$.  For the former, we determine
momentum $q$ by requiring the total current to be zero.
From $\mathbf{j}=(\nabla_{\mathbf{x}}-\nabla_{\mathbf{x^{'}}}) |_{x\to
x^{'}}\mathbf{G^{\prime}}(x,x^{'})=\sum_{K}[\mathbf{k}G_{\uparrow}(K)+\mathbf{k}G_{\downarrow}(K)]$ we get
\begin{widetext}
\begin{eqnarray}
\mathbf{j}
&=&\sum_{\mathbf{k}}\mathbf{k}[f(E_{1,\uparrow})u_{\mathbf{k}}^{2}
+f(-E_{1,\downarrow})v_{\mathbf{k}}^{2}]-\sum_{\mathbf{k}}\mathbf{(k-q)}[f(E_{1,\downarrow})u_{\mathbf{k}}^{2}+f(-E_{1,\uparrow})v_{\mathbf{k}}^{2}]\nonumber \\
&=&\sum_{\mathbf{k}}\Big[\frac{\mathbf{q}}{2}
\Big(1-\frac{\xi_{kq}}{E_{kq}}\Big)+\frac{\mathbf{q}}{2}[f(E_{1,\uparrow})+f(E_{1,\downarrow})]\frac{\xi_{kq}}{E_{kq}}+(\mathbf{k-}\frac{\mathbf{q}}{2})[f(E_{1,\uparrow})-f(E_{1,\downarrow})]\Big] \nonumber \\
&=&0\nonumber
\end{eqnarray}
\end{widetext}

We can take $\mathbf{q}$ as the $z$ direction, and all the above
equations need to be integrated over the angular variable $\theta$ .

The pair susceptibility, even for the simpler LOFF state is reasonably
complex so we present it here for completeness.
$\chi(P)=\frac{1}{2}\big[\chi_{\uparrow\downarrow}(P)
+\chi_{\downarrow\uparrow}(P)\big]$ is
\begin{widetext}
\begin{equation}\label{eq:LOFF1chi}
\chi(P)=\frac{1}{2}\sum_{\mathbf{k}}\Big[
u_{k}^{2}\frac{f(E_{1,\uparrow})+f(E_{2,\downarrow})
+f(\xi_{p-k,\uparrow})+f(\xi_{p-k,\downarrow})-2}
{\Omega-\xi_{p-k}-(E_{kq}+(\epsilon_k-\epsilon_{k-q})/2)}+v_{k}^{2}\frac{f(\xi_{p-k,\uparrow})+f(\xi_{p-k,\downarrow})
-f(E_{2,\uparrow})-f(E_{1,\downarrow})}
{\Omega-\xi_{p-k}+(E_{kq}-(\epsilon_k-\epsilon_{k-q})/2)}\Big].
\end{equation}
\end{widetext}

Next, we characterize the pseudogap contributions which enter via
$\Delta_{pg}^{2}=Z^{-1}\sum_{\mathbf{p}}b(\Omega_{p})$.
Here $Z$ and $\frac{1}{2M^*}$ are determined as follows.

\begin{eqnarray}
Z&=&\frac{\partial t^{-1}}{\partial\Omega}\Big|_{\Omega=0,\mathbf{p=q}} \nonumber \\
&=&\frac{1}{2\Delta^{2}}\Big\{n
-\sum_{\mathbf{k}}[f(\xi_{k-q\uparrow})+f(\xi_{k-q\downarrow})]\Big\}.
\end{eqnarray}
Here $n$ is the total density.
Note that the linear derivative vanishes, as was established in 
Section \ref{sec:IIB}. Thus we turn next to the quadratic term.

The coefficient of this second order term corresponds to the inverse
pair mass, $1/2M^*$.  This is given by


\begin{widetext}
\begin{eqnarray}
\frac{1}{2M^*}&=&-\frac{1}{6Z}\frac{\partial^{2}\chi(0,\mathbf{p})}{\partial p^{2}}\Big|_{\mathbf{p=q}} \nonumber \\
&=&-\frac{1}{12mZ\Delta^{2}}\sum_{\mathbf{k}}\Big\{
2\bar{f}(\xi_{\mathbf{k-q}})
\Big[3+\frac{4\xi_{kq}(\mathbf{k-q})^2}{m\Delta^{2}}\Big]-[f(E_{1,\uparrow})+f(E_{1,\downarrow})]\Big[\frac{3\xi_{kq}}{E_{kq}}
+\frac{2[(\mathbf{k-q})^2+q^2](E_{kq}^{2}+\xi_{kq}^{2})}{mE_{kq}\Delta^{2}}\Big] \nonumber \\
& &{}+\frac{2[f(E_{1,\uparrow})-f(E_{1,\downarrow})]}{m\Delta^2}[(\mathbf{k-q})^2-q^2]\xi_{kq}
+\frac{8}{m}\bar{f}^{\prime}(\xi_{\mathbf{k-q}})(\mathbf{k-q})^2-3\Big(1-\frac{\xi_{kq}}{E_{kq}}\Big) \nonumber \\
& &{}+\frac{2E_{kq}}{m\Delta^{2}}\Big(1-\frac{\xi_{kq}}{E_{kq}}\Big)^{2}(\mathbf{k-q})^2\Big\}
\end{eqnarray}
\end{widetext}
 
\section{Additional Results for the Two plane Wave LOFF state}
\label{App:LOFF2}

We begin with the results that relate to the simplest mean field-based
scheme discussed in Section \ref{sec:IIIA}.  We then proceed to the
slightly different scheme based on a resummation of the LO diagrams.

The pair susceptibility is given by
\begin{widetext}
\begin{eqnarray}\label{eq:LOFF2symchi}
\chi(P)&=&\frac{1}{2}\sum_{K}[G_{0\uparrow}(P-K)
G_{\downarrow}(K)+G_{0\downarrow}(P-K)G_{\uparrow}(K)] \nonumber \\
&=&\frac{1}{2}T\sum_{n}\sum_{\mathbf{k},\sigma}\frac{1}{i\Omega-i\omega_{n}-
\xi_{\mathbf{p-k}\sigma}}\,\frac{(i\omega_{n}+
\xi_{\mathbf{k+q}\sigma})(i\omega_{n}+\xi_{\mathbf{k-q}\sigma})}
{(i\omega_{n}-E_{1\bar{\sigma}})(i\omega_{n}-E_{2\bar{\sigma}})
(i\omega_{n}-E_{3\bar{\sigma}})}
\end{eqnarray}

From the pair susceptibility one may obtain the dispersion relation for
the non-condensed pairs.  This is important for introducing pseudogap
effects and treating $T \neq 0$.


The explicit forms of the coefficients $Z$ and $1/2M^*$ are
\begin{eqnarray}
Z
&=&-\frac{1}{2}\sum_{\mathbf{k}}\Big\{\frac{\frac{2}{m}\mathbf{k\cdot q}[f(\xi_{\mathbf{k-q}\uparrow})+f(\xi_{\mathbf{k-q}\downarrow})-2]}{(E_{1}+\mathcal{E}_{kq}-\frac{\mathbf{k\cdot q}}{m})(E_{2}+\mathcal{E}_{kq}-\frac{\mathbf{k\cdot q}}{m})(E_{3}+\mathcal{E}_{kq}-\frac{\mathbf{k\cdot q}}{m})}+\frac{f(E_{1\uparrow})+f(E_{1\downarrow})}{(E_{1}-E_{2})(E_{1}-E_{3})}\,\frac{E_{1}+\mathcal{E}_{kq}+\frac{\mathbf{k\cdot q}}{m}}{E_{1}+\mathcal{E}_{kq}-\frac{\mathbf{k\cdot q}}{m}} \nonumber \\
& &{}+\frac{f(E_{2\uparrow})+f(E_{2\downarrow})}{(E_{2}-E_{1})(E_{2}-E_{3})}\,\frac{E_{2}+\mathcal{E}_{kq}+\frac{\mathbf{k\cdot q}}{m}}{E_{2}+\mathcal{E}_{kq}-\frac{\mathbf{k\cdot q}}{m}}+\frac{f(E_{3\uparrow})+f(E_{3\downarrow})}{(E_{3}-E_{1})(E_{3}-E_{1})}\,\frac{E_{3}+\mathcal{E}_{kq}+\frac{\mathbf{k\cdot q}}{m}}{E_{3}+\mathcal{E}_{kq}-\frac{\mathbf{k\cdot q}}{m}} \Big\}
\end{eqnarray}
and
\begin{eqnarray}
\frac{1}{2M^*}
&=&\frac{1}{24mZ}\sum_{\mathbf{k}}\Big\{\frac{f(\xi_{\mathbf{k-q}\uparrow})+f(\xi_{\mathbf{k-q}\downarrow})-2}{(E_{1}+\mathcal{E}_{kq}-\frac{\mathbf{k\cdot q}}{m})(E_{2}+\mathcal{E}_{kq}-\frac{\mathbf{k\cdot q}}{m})(E_{3}+\mathcal{E}_{kq}-\frac{\mathbf{k\cdot q}}{m})}\,\Big[-\frac{12}{m}(\mathbf{k\cdot q})+\frac{4}{m}(\xi_{\mathbf{k-q}}+\mu) \nonumber \\
& &{}+\frac{8}{m^{2}}(\mathbf{k\cdot q})(\xi_{\mathbf{k-q}}+\mu)\Big(\frac{1}{E_{1}+\mathcal{E}_{kq}-\frac{\mathbf{k\cdot q}}{m}}+\frac{1}{E_{2}+\mathcal{E}_{kq}-\frac{\mathbf{k\cdot q}}{m}}+\frac{1}{E_{3}+\mathcal{E}_{kq}-\frac{\mathbf{k\cdot q}}{m}} \Big)\Big] \nonumber \\
& &{}-\frac{\frac{8}{m^{2}}(\mathbf{k\cdot q})(\xi_{\mathbf{k-q}}+\mu)[f^{\prime}(\xi_{\mathbf{k-q}\uparrow})+f^{\prime}(\xi_{\mathbf{k-q}\downarrow})]}{(E_{1}+\mathcal{E}_{kq}-\frac{\mathbf{k\cdot q}}{m})(E_{2}+\mathcal{E}_{kq}-\frac{\mathbf{k\cdot q}}{m})(E_{3}+\mathcal{E}_{kq}-\frac{\mathbf{k\cdot q}}{m})} \nonumber \\
& &{}+\frac{f(E_{1\uparrow})+f(E_{1\downarrow})}{(E_{1}-E_{2})(E_{1}-E_{3})}\,\Big[\frac{\frac{4}{m}(\xi_{\mathbf{k-q}}+\mu)(E_{1}+\mathcal{E}_{kq}+\frac{\mathbf{k\cdot q}}{m})}{(E_{1}+\mathcal{E}_{kq}-\frac{\mathbf{k\cdot q}}{m})^{2}}-6\frac{E_{1}+\mathcal{E}_{kq}+\frac{\mathbf{k\cdot q}}{m}}{E_{1}+\mathcal{E}_{kq}-\frac{\mathbf{k\cdot q}}{m}}\Big] \nonumber \\
& &{}+\frac{f(E_{2\uparrow})+f(E_{2\downarrow})}{(E_{2}-E_{1})(E_{2}-E_{3})}\,\Big[\frac{\frac{4}{m}(\xi_{\mathbf{k-q}}+\mu)(E_{2}+\mathcal{E}_{kq}+\frac{\mathbf{k\cdot q}}{m})}{(E_{2}+\mathcal{E}_{kq}-\frac{\mathbf{k\cdot q}}{m})^{2}}-6\frac{E_{2}+\mathcal{E}_{kq}+\frac{\mathbf{k\cdot q}}{m}}{E_{2}+\mathcal{E}_{kq}-\frac{\mathbf{k\cdot q}}{m}}\Big] \nonumber \\
& &{}+\frac{f(E_{2\uparrow})+f(E_{2\downarrow})}{(E_{2}-E_{1})(E_{2}-E_{3})}\,\Big[\frac{\frac{4}{m}(\xi_{\mathbf{k-q}}+\mu)(E_{3}+\mathcal{E}_{kq}+\frac{\mathbf{k\cdot q}}{m})}{(E_{3}+\mathcal{E}_{kq}-\frac{\mathbf{k\cdot q}}{m})^{2}}-6\frac{E_{3}+\mathcal{E}_{kq}+\frac{\mathbf{k\cdot q}}{m}}{E_{3}+\mathcal{E}_{kq}-\frac{\mathbf{k\cdot q}}{m}}\Big] \Big\}
\end{eqnarray}

For the LO-based pair susceptibility we have
%
\begin{eqnarray}
\chi^{LO} (P)
&=&\sum_{K}[G_{0\uparrow}(P-K)
G_{\downarrow}(K)+G_{0\downarrow}(P-K)G_{\uparrow}(K)]-\frac{1}{2}\sum_{K}[G_{0\uparrow}(P-K)
G^{(1)}_{\downarrow}(K)+G_{0\downarrow}(P-K)G^{(1)}_{\uparrow}(K)] 
\nonumber \\
&=&T\sum_{n}\sum_{\mathbf{k},\sigma}\Big\{\frac{1}{i\Omega_l-i\omega_{n}-
\xi_{\mathbf{p-k}\sigma}}\,\frac{(i\omega_{n}+
\xi_{\mathbf{k+q}\sigma})(i\omega_{n}+\xi_{\mathbf{k-q}\sigma})}
{(i\omega_{n}-E_{1\bar{\sigma}})(i\omega_{n}-E_{2\bar{\sigma}})
(i\omega_{n}-E_{3\bar{\sigma}})}\Big\} \nonumber \\
& &{}-\frac{1}{2}T\sum_{n}\sum_{\mathbf{k}}\Big\{\frac{1}{i\Omega_l-i\omega_{n}
-\xi_{\mathbf{p-k}\uparrow}}\,\frac{i\omega_{n}+
\xi_{\mathbf{k-q}\uparrow}}{(i\omega_{n}-E^{(1)}_{2,\downarrow})
(i\omega_{n}+E^{(1)}_{2,\uparrow})} \nonumber \\
& &{}+\frac{1}{i\Omega_l-i\omega_{n}-\xi_{\mathbf{p-k}\downarrow}}
\,\frac{i\omega_{n}+\xi_{\mathbf{k-q}\downarrow}}
{(i\omega_{n}-E^{(1)}_{1,\uparrow})(i\omega_{n}
+E^{(1)}_{1,\downarrow})}\Big\}
\end{eqnarray}

 The momentum $\mathbf{q}$ is determined by minimizing
 $\chi(0,\mathbf{p})$, i.e., $\frac{\partial
 \chi(0,\mathbf{p})}{\partial\mathbf{p}}=0$ at
 $\mathbf{p}=\mathbf{q}$. This gives the \textbf{momentum equation}
\begin{eqnarray}
0&=&T\sum_{n}\sum_{\mathbf{k},\sigma}\Big(\frac{\mathbf{k-q}}{m}\Big)
\frac{i\omega_{n}+\xi_{\mathbf{k+q}\sigma}}
{i\omega_{n}+\xi_{\mathbf{k-q}\sigma}}\,\frac{1}
{(i\omega_{n}-E_{1\bar{\sigma}})(i\omega_{n}-E_{2\bar{\sigma}})
(i\omega_{n}-E_{3\bar{\sigma}})} \nonumber \\
& &-\frac{1}{2}T\sum_{n}\sum_{\mathbf{k}}\Big(\frac{\mathbf{k-q}}{m}\Big)
\left[\frac{1}{i\omega_{n}+\xi_{\mathbf{k-q}\uparrow}}
\,\frac{1}{(i\omega_{n}-E^{(1)}_{2,\downarrow})
(i\omega_{n}+E^{(1)}_{2,\uparrow})}+\frac{1}{i\omega_{n}+\xi_{\mathbf{k-q}\downarrow}}
\,\frac{1}{(i\omega_{n}-E^{(1)}_{1,\uparrow})
(i\omega_{n}+E^{(1)}_{1,\downarrow})} \right] \nonumber \\
&=&\sum_{\mathbf{k}}\Big(\frac{\mathbf{k-q}}{m}\Big)\Big\{\frac{\frac{2}{m}(\mathbf{k\cdot q})[f(\xi_{\mathbf{k-q}\uparrow})+f(\xi_{\mathbf{k-q}\downarrow})-2]}{(E_{1}+\mathcal{E}_{kq}-\frac{\mathbf{k\cdot q}}{m})(E_{2}+\mathcal{E}_{kq}-\frac{\mathbf{k\cdot q}}{m})(E_{3}+\mathcal{E}_{kq}-\frac{\mathbf{k\cdot q}}{m})} \nonumber \\
& &{}+\frac{f(E_{1\uparrow})+f(E_{1\downarrow})}{(E_{1}-E_{2})(E_{1}-E_{3})}\,\frac{E_{1}+\mathcal{E}_{kq}+\frac{\mathbf{k\cdot q}}{m}}{E_{1}+\mathcal{E}_{kq}-\frac{\mathbf{k\cdot q}}{m}}+\frac{f(E_{2\uparrow})+f(E_{2\downarrow})}{(E_{2}-E_{1})(E_{2}-E_{3})}\,\frac{E_{2}+\mathcal{E}_{kq}+\frac{\mathbf{k\cdot q}}{m}}{E_{2}+\mathcal{E}_{kq}-\frac{\mathbf{k\cdot q}}{m}} \nonumber \\
& &{}+\frac{f(E_{3\uparrow})+f(E_{3\downarrow})}{(E_{3}-E_{1})(E_{3}-E_{1})}\,\frac{E_{3}+\mathcal{E}_{kq}+\frac{\mathbf{k\cdot q}}{m}}{E_{3}+\mathcal{E}_{kq}-\frac{\mathbf{k\cdot q}}{m}}\Big\} \nonumber \\
& &{}-\frac{1}{2\Delta^{2}}\sum_{\mathbf{k}}\Big\{\frac{\mathbf{q}}{2}\Big(1-\frac{\xi^{(1)}_{\mathbf{kq}}}{E^{(1)}_{kq}}\Big)+\frac{\mathbf{q}}{2}\frac{\xi^{(1)}_{\mathbf{kq}}}{E^{(1)}_{kq}}[f(E^{(1)}_{1,\uparrow})+f(E^{(1)}_{1,\downarrow})]+(\mathbf{k-}\frac{\mathbf{q}}{2})[f(E^{(1)}_{1,\uparrow})-f(E^{(1)}_{1,\downarrow})]\Big\}.
\end{eqnarray}

The coefficients in the pseudogap dispersion are
\begin{eqnarray}
Z
&=&-\sum_{\mathbf{k}}\left\{\frac{\frac{2}{m}(\mathbf{k\cdot q})[f(\xi_{\mathbf{k-q}\uparrow})+f(\xi_{\mathbf{k-q}\downarrow})-2]}{(E_{1}+\mathcal{E}_{kq}-\frac{\mathbf{k\cdot q}}{m})(E_{2}+\mathcal{E}_{kq}-\frac{\mathbf{k\cdot q}}{m})(E_{3}+\mathcal{E}_{kq}-\frac{\mathbf{k\cdot q}}{m})}+\frac{f(E_{1\uparrow})+f(E_{1\downarrow})}{(E_{1}-E_{2})(E_{1}-E_{3})}\,\frac{E_{1}+\mathcal{E}_{kq}+\frac{\mathbf{k\cdot q}}{m}}{E_{1}+\mathcal{E}_{kq}-\frac{\mathbf{k\cdot q}}{m}}\right. \nonumber \\
& &{}+\frac{f(E_{2\uparrow})+f(E_{2\downarrow})}{(E_{2}-E_{1})(E_{2}-E_{3})}\,\frac{E_{2}+\mathcal{E}_{kq}+\frac{\mathbf{k\cdot q}}{m}}{E_{2}+\mathcal{E}_{kq}-\frac{\mathbf{k\cdot q}}{m}}+\frac{f(E_{3\uparrow})+f(E_{3\downarrow})}{(E_{3}-E_{1})(E_{3}-E_{1})}\,\frac{E_{3}+\mathcal{E}_{kq}+\frac{\mathbf{k\cdot q}}{m}}{E_{3}+\mathcal{E}_{kq}-\frac{\mathbf{k\cdot q}}{m}}  \nonumber \\
& &{}\left. + \frac{1}{2\Delta^{2}}\bigg[[1-f(\xi_{\mathbf{k-q}\uparrow})-f(\xi_{\mathbf{k-q}\downarrow})]
-\frac{\xi^{(1)}_{\mathbf{k\cdot q}}}{E^{(1)}_{kq}}[1-f(E^{(1)}_{1,\uparrow})-f(E^{(1)}_{1,\downarrow})]\bigg]\right\}
\end{eqnarray}
and
\begin{eqnarray}
\frac{1}{2M^*}
&=&\frac{1}{12mZ}\sum_{\mathbf{k}}\Big\{\frac{f(\xi_{\mathbf{k-q}\uparrow})+f(\xi_{\mathbf{k-q}\downarrow})-2}{(E_{1}+\mathcal{E}_{kq}-\frac{\mathbf{k\cdot q}}{m})(E_{2}+\mathcal{E}_{kq}-\frac{\mathbf{k\cdot q}}{m})(E_{3}+\mathcal{E}_{kq}-\frac{\mathbf{k\cdot q}}{m})}\,\Big[-\frac{12}{m}(\mathbf{k\cdot q})+\frac{4}{m}(\xi_{\mathbf{k-q}}+\mu) \nonumber \\
& &{}+\frac{8}{m^{2}}(\mathbf{k\cdot q})(\xi_{\mathbf{k-q}}+\mu)\Big(\frac{1}{E_{1}+\mathcal{E}_{kq}-\frac{\mathbf{k\cdot q}}{m}}+\frac{1}{E_{2}+\mathcal{E}_{kq}-\frac{\mathbf{k\cdot q}}{m}}+\frac{1}{E_{3}+\mathcal{E}_{kq}-\frac{\mathbf{k\cdot q}}{m}} \Big)\Big] \nonumber \\
& &{}-\frac{\frac{8}{m^{2}}(\mathbf{k\cdot q})(\xi_{\mathbf{k-q}}+\mu)[f^{\prime}(\xi_{\mathbf{k-q}\uparrow})+f^{\prime}(\xi_{\mathbf{k-q}\downarrow})]}{(E_{1}+\mathcal{E}_{kq}-\frac{\mathbf{k\cdot q}}{m})(E_{2}+\mathcal{E}_{kq}-\frac{\mathbf{k\cdot q}}{m})(E_{3}+\mathcal{E}_{kq}-\frac{\mathbf{k\cdot q}}{m})} \nonumber \\
& &{}+\frac{f(E_{1\uparrow})+f(E_{1\downarrow})}{(E_{1}-E_{2})(E_{1}-E_{3})}\,\Big[\frac{\frac{4}{m}(\xi_{\mathbf{k-q}}+\mu)(E_{1}+\mathcal{E}_{kq}+\frac{\mathbf{k\cdot q}}{m})}{(E_{1}+\mathcal{E}_{kq}-\frac{\mathbf{k\cdot q}}{m})^{2}}-6\frac{E_{1}+\mathcal{E}_{kq}+\frac{\mathbf{k\cdot q}}{m}}{E_{1}+\mathcal{E}_{kq}-\frac{\mathbf{k\cdot q}}{m}}\Big] \nonumber \\
& &{}+\frac{f(E_{2\uparrow})+f(E_{2\downarrow})}{(E_{2}-E_{1})(E_{2}-E_{3})}\,\Big[\frac{\frac{4}{m}(\xi_{\mathbf{k-q}}+\mu)(E_{2}+\mathcal{E}_{kq}+\frac{\mathbf{k\cdot q}}{m})}{(E_{2}+\mathcal{E}_{kq}-\frac{\mathbf{k\cdot q}}{m})^{2}}-6\frac{E_{2}+\mathcal{E}_{kq}+\frac{\mathbf{k\cdot q}}{m}}{E_{2}+\mathcal{E}_{kq}-\frac{\mathbf{k\cdot q}}{m}}\Big] \nonumber \\
& &{}+\frac{f(E_{2\uparrow})+f(E_{2\downarrow})}{(E_{2}-E_{1})(E_{2}-E_{3})}\,\Big[\frac{\frac{4}{m}(\xi_{\mathbf{k-q}}+\mu)(E_{3}+\mathcal{E}_{kq}+\frac{\mathbf{k\cdot q}}{m})}{(E_{3}+\mathcal{E}_{kq}-\frac{\mathbf{k\cdot q}}{m})^{2}}-6\frac{E_{3}+\mathcal{E}_{kq}+\frac{\mathbf{k\cdot q}}{m}}{E_{3}+\mathcal{E}_{kq}-\frac{\mathbf{k\cdot q}}{m}}\Big] \Big\} \nonumber \\
& &{}-\frac{1}{12mZ\Delta_{1}^{2}}\sum_{\mathbf{k}}\Big\{
[f(\xi_{\mathbf{k-q}\uparrow})+f(\xi_{\mathbf{k-q}\downarrow})]
\Big[3+\frac{4\xi^{(1)}_{\mathbf{kq}}(\mathbf{k-q})^2}{m\Delta_{1}^{2}}\Big] \nonumber \\
& &{}-[f(E^{(1)}_{1,\uparrow})+f(E^{(1)}_{1,\downarrow})]\Big[\frac{3\xi^{(1)}_{\mathbf{kq}}}{E^{(1)}_{kq}}
+\frac{2[(\mathbf{k-q})^2+q^2]({E^{(1)}_{kq}}^{2}+{\xi^{(1)}_{\mathbf{kq}}}^{2})}{mE^{(1)}_{kq}\Delta_{1}^{2}}\Big]+\frac{2[f(E^{(1)}_{1,\uparrow})-f(E^{(1)}_{1,\downarrow})]}{m\Delta_{1}^2}[(\mathbf{k-q})^2-q^2]\xi^{(1)}_{\mathbf{kq}} \nonumber \\
& &{}+\frac{2}{m}[f^{\prime}(\xi_{\mathbf{k-q}\uparrow})+f^{\prime}(\xi_{\mathbf{k-q}\downarrow})](\mathbf{k-q})^2-3\Big(1-\frac{\xi^{(1)}_{\mathbf{kq}}}{E^{(1)}_{kq}}\Big)
+\frac{2E^{(1)}_{kq}}{m\Delta_{1}^{2}}\Big(1-\frac{\xi^{(1)}_{\mathbf{kq}}}{E^{(1)}_{kq}}\Big)^{2}(\mathbf{k-q})^2\Big\}.
\end{eqnarray}
Here we define ${u_{kq}^{(1)}}^{2}=\frac{1}{2}\Big(1+\frac{\xi^{(1)}_{\mathbf{kq}}}{E^{(1)}_{kq}}\Big)$ and ${v_{kq}^{(1)}}^{2}=\frac{1}{2}\Big(1-\frac{\xi^{(1)}_{\mathbf{kq}}}{E^{(1)}_{kq}}\Big)$, where $\xi^{(1)}_{\mathbf{kq}}=\mathcal{E}_{kq}-\mathbf{k\cdot q}/2m$.

\end{widetext}


\bibliographystyle{apsrev}


\end{document}